\documentclass[journal, romanappendices, final]{IEEEtran}
\usepackage{latexsym,amsfonts,amssymb,amsthm, setspace, verbatim,cite,mathrsfs,graphicx,bm,stfloats, hyperref, tikz,xcolor, bm, algorithmic, algorithm, enumerate, paralist, siunitx, flushend}
\usepackage[cmex10]{amsmath}
\newcommand{\be}{\begin{equation}}
\newcommand{\ee}{\end{equation}}
\newcommand{\ben}{\begin{equation*}}
\newcommand{\een}{\end{equation*}}
\newcommand{\mc}{\mathcal}
\newcommand{\mbf}{\mathbf}
\newcommand{\e}{\epsilon}

\newcommand{\wh}[1]{\widehat{#1}}

\newcommand{\abs}[1]{\lvert#1\rvert}
\newcommand{\norm}[1]{\lVert#1\rVert}
\newcommand{\expec}{\mathbb{E}}

\newcommand{\snr}{\textsf{snr}}
\newcommand{\Var}{\mathrm{Var}}
\newcommand{\Prob}{\mathbb{P}}
\newcommand{\C}{\mc{C}}
\newcommand{\Eb}{\mc{E}_{\text{ber}}}
\newcommand{\Es}{\mc{E}_{\text{sec}}}
\newcommand{\Ec}{\mc{E}_{\text{cw}}}
\newcommand{\algorithmicbreak}{\textbf{break}}
\newcommand{\BREAK}{\STATE \algorithmicbreak}
\newtheorem{lem}{Lemma}

\newtheorem{prop}{Proposition}

\begin{document}
\title{Techniques for Improving the Finite Length  Performance  of   Sparse Superposition Codes}
\author{Adam Greig,~\IEEEmembership{Student Member,~IEEE,}
and Ramji Venkataramanan,~\IEEEmembership{Senior Member,~IEEE}
\thanks{A.~Greig and R.~Venkataramanan are with Department of Engineering, University of Cambridge, Cambridge CB2 1PZ, UK (e-mails: ag611@cam.ac.uk, \, rv285@cam.ac.uk).}
\thanks{This work was supported in part by EPSRC Grant EP/N013999/1, and by an EPSRC Doctoral Training Award.}
\vspace{-10pt}}
\maketitle

\begin{abstract}
Sparse superposition codes are a recent class of codes introduced by Barron and Joseph for efficient communication over the AWGN channel. With an appropriate power allocation, these codes have been shown to be asymptotically capacity-achieving with computationally feasible decoding. However, a direct implementation of the  capacity-achieving construction  does not give good finite length error performance. In this paper, we consider sparse superposition codes with approximate message passing (AMP) decoding, and  describe a variety of techniques to improve their finite length performance. These include an iterative algorithm  for SPARC power allocation, guidelines for choosing codebook parameters, and estimating a critical decoding parameter online instead of pre-computation.  We also show how partial outer codes can be used in conjunction with AMP decoding to obtain a steep waterfall in the error performance curves.  We compare the error performance of AMP-decoded sparse superposition codes  with coded modulation using LDPC codes from the WiMAX standard.
\end{abstract}


\section{Introduction}
\label{sec:intro}
\IEEEPARstart{W}e consider communication over the memoryless additive white Gaussian noise (AWGN) channel given by  \[ y = x + w, \] 
where the channel output $y$ is the sum of the channel input $x$  and independent zero-mean Gaussian noise $w$ of variance $\sigma^2$. There is an average power constraint $P$ on the input, so a length-$n$ codeword $(x_1, \ldots, x_n)$ has to satisfy $\frac{1}{n}\sum_{i=1}^{n}x_i^2 \le P$.  The goal is to build computationally efficient codes that have low probability
of decoding error at rates close to the AWGN channel capacity $\mc{C} = \frac{1}{2}\log(1+ \snr)$. Here $\snr$  denotes the signal-to-noise ratio $P/\sigma^2$.

Though it is well known that Shannon-style i.i.d. Gaussian codebooks can achieve very low probability of error at rates approaching the AWGN capacity \cite{gallager68book}, this approach has been largely avoided in practice due to the high decoding complexity of unstructured Gaussian codes.  Current state of the art approaches for the AWGN channel such as coded modulation \cite{bicmAlbert, bocherer15}  typically involve separate {coding} and  {modulation} steps. In this approach,  a binary error-correcting code such as an LDPC or turbo code is first used to generate a binary codeword from the information bits; the code bits are then modulated with a standard scheme such as quadrature amplitude modulation. Though these schemes have good empirical performance, they have not been proven to be capacity-achieving for the AWGN channel.

Sparse Superposition Codes or Sparse Regression Codes (SPARCs) were recently proposed by Barron and Joseph \cite{Antony2012, AntonyFast} for efficient communication over the AWGN channel. In \cite{AntonyFast}, they introduced an efficient decoding algorithm called ``adaptive successive decoding" and showed that it achieved near-exponential decay of  error probability (with growing block length), for any fixed rate $R < \mc{C}$. Subsequently, an adaptive soft-decision successive decoder was proposed in \cite{BarronC12,choBarron13}, and  Approximate Message Passing (AMP) decoders were proposed in \cite{barbier2014replica, BarbSK2015, BarbKrz15, Rush2017}.
The adaptive soft-decision decoder in \cite{choBarron13} as well as the AMP decoder in \cite{Rush2017} were proven to be asymptotically capacity-achieving, and have superior finite length performance compared to the original adaptive successive decoder of \cite{AntonyFast}.

The above results mainly focused on characterizing the error performance of SPARCs in the limit of large block length.
In this work, we describe a number of code design techniques for improved \emph{finite length} error performance.  Throughout the paper, we focus on AMP decoding due to its ease of implementation. However, many of the code design ideas can also be applied to the adaptive soft-decision successive decoder in \cite{BarronC12,choBarron13}. A hardware implementation  of the AMP decoder was recently reported in 
\cite{condo2015sparse,condoGross17}. We expect that the techniques proposed in this paper can be used to reduce the complexity and optimize the decoding performance in such implementations.

In the remainder of this section, we  briefly review the SPARC construction and the AMP decoder from \cite{Rush2017}, and then list the main contributions of this paper.  A word about notation before we proceed. Throughout the paper, we use $\log$ to denote logarithms with base $2$, and $\ln$ to denote natural logarithms. For a positive integer $N$, we use $[N]$ to denote the set $\{1, \ldots, N\}$. The transpose of a matrix $A$ is denoted by $A^*$, and the indicator function of an event $\mc{E}$  by $\mbf{1}\{ \mc{E} \}$.

\subsection{The sparse superposition code}

\begin{figure}[t]
\centering
\includegraphics[width=\columnwidth]{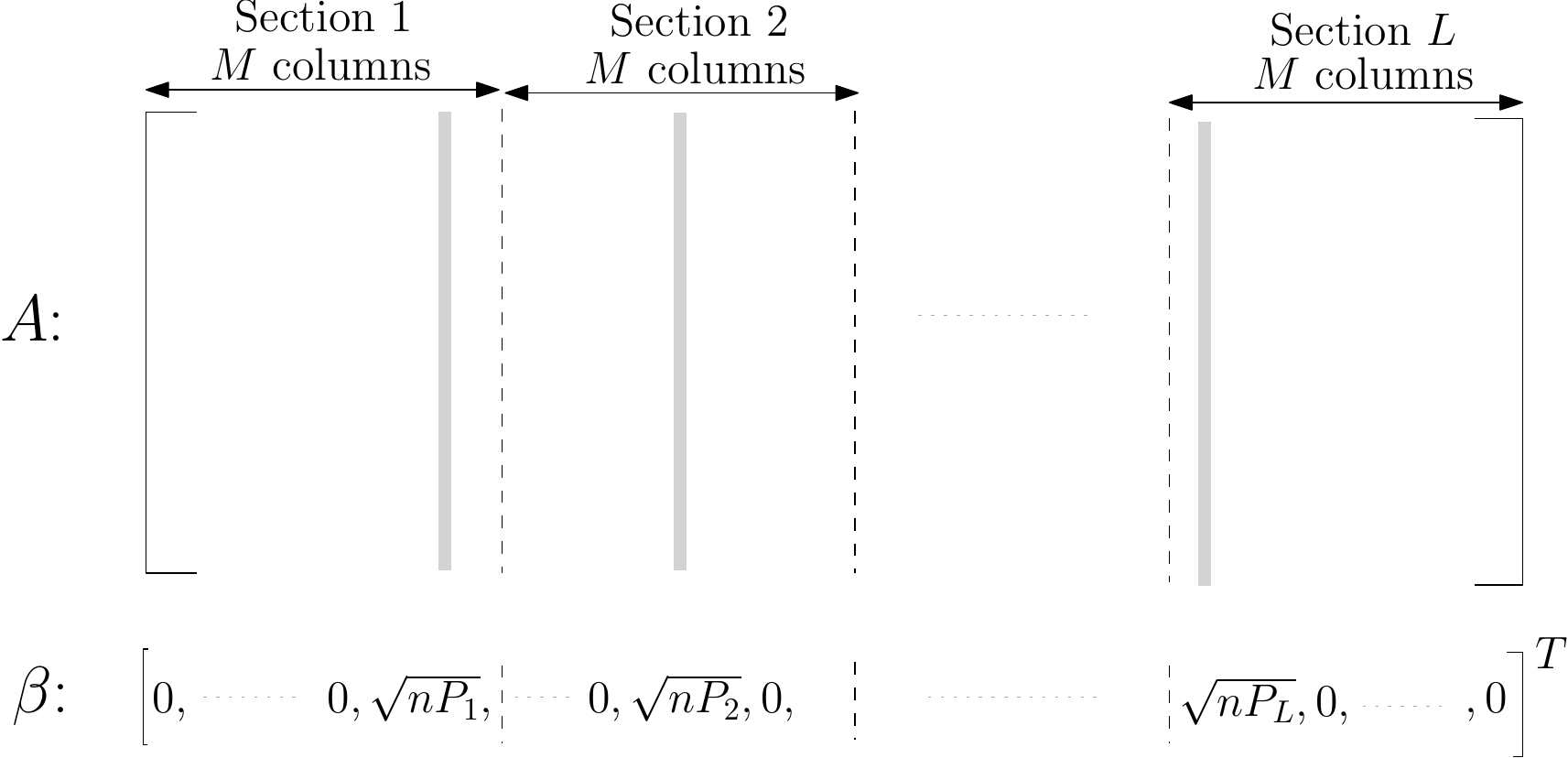}
\caption{\small{$A$ is the $n\times LM$ design matrix, $\beta$ is an $ML\times 1$ sparse vector with one non-zero in each of the $L$ sections. The length-$n$ codeword is $A\beta$. The message determines the locations of the non-zeros in $\beta$, while
$P_1,\ldots,P_L$ are fixed a priori.}}%
\label{fig:sparse_design_matrix}
\vspace{-5pt}
\end{figure}

A SPARC is defined in terms of a  design matrix $A$ of dimension $n \times ML$. Here $n$ is the block length, and $M, L$  are integers which are specified below in terms of $n$ and the rate $R$.  As shown in Fig. \ref{fig:sparse_design_matrix}, the design matrix $A$ has $L$ sections with $M$ columns each.  In the original construction  of \cite{Antony2012, AntonyFast}  and in the theoretical analysis in \cite{BarronC12,choBarron13,Rush2017,Rush2017b}, the entries of $A$ are assumed to be i.i.d. Gaussian $\sim \mc{N}(0,1/n)$. For our empirical results, we use a random Hadamard-based construction for $A$ that leads to significantly lower encoding and decoding complexity \cite{BarbSK2015,BarbKrz15,Rush2017}.

Codewords are constructed as sparse linear combinations of the columns of $A$. In particular, a codeword is of the form $A\beta$, where $\beta = (\beta_1, \ldots, \beta_{ML})^*$ is a length $ML$ column
vector  with the property that there is exactly one non-zero $\beta_j$ for the section $1\le j \le M$, one non-zero $\beta_j$ for the section $M+1\le j \le 2M$, and so forth. The non-zero value of $\beta$
in each section $\ell$ is set to $\sqrt{nP_\ell}$, where $P_1,\ldots,P_L$ are pre-specified positive constants that satisfy $\sum_{\ell=1}^L
P_\ell=P$, the average symbol power allowed.  

Both $A$ and the \emph{power allocation}  $\{P_1, \ldots, P_L \}$ are known to both the encoder and decoder in advance. The choice of {power allocation} plays a crucial role in determining the error performance of the decoder. Without loss of generality, we will assume that the power allocation is non-increasing across sections. Two examples of power allocation are: 
\begin{itemize}
\item Flat power allocation, where $P_\ell = \frac{P}{L}$ for all $\ell$. This choice was used  in \cite{Antony2012} to analyze the error performance with optimal (least-squares) decoding.

\item Exponentially decaying power allocation, where $P_\ell \propto 2^{-{2\mc{C} \ell /L}}$. This choice was used for the asymptotically capacity-achieving decoders proposed in \cite{AntonyFast,choBarron13,Rush2017}. 
\end{itemize}
At finite block lengths both these power allocations could be far from optimal and  lead to poor decoding performance. One of the main contributions of this paper is an algorithm to determine a good power allocation for the finite-length AMP decoder based only on $R$, $P$, $\sigma^2$.

\emph{Rate}: As each of the $L$ sections contains $M$ columns, the total number of
codewords is $M^L$. With the block length being $n$, the rate of the code is given by 
\be R = \frac{\log (M^L)}{n} = \frac{L \log M}{n}. \label{eq:Rdef} \ee
In other words, a SPARC codeword corresponding to $L \log M$ input bits is transmitted in $n$ channel uses.

\emph{Encoding}:  The input bitstream is split into chunks of $\log M$ bits.  A chunk of $\log M$ input bits can be used to  index the location of the non-zero entry in one section of $\beta$. Hence $L$ successive chunks determine the message vector $\beta$, with the $\ell$th chunk of $\log M$ input  bits determining the non-zero location in section $\ell$, for $1 \leq \ell \leq L$.  

\emph{Approximate Message Passing (AMP) decoder}: The AMP decoder produces iteratively refined estimates of the message vector, denoted by $\beta^1, \beta^2, \ldots, \beta^T$, where $T$ is the (pre-specified) number of iterations. Starting with  $\beta^0=0$,  for $t=0,1,\ldots, T-1$ the AMP decoder generates
\begin{align}
    z^t & = y - A\beta^t + \frac{z^{t-1}}{\tau^2_{t-1}}\left(P - \frac{\norm{\beta^t}^2}{n}\right), \label{eq:z_update} \\ 
\beta_i^{t+1} & = \eta_i^t(\beta^t + A^*z^t), \label{eq:beta_update}   
\end{align}
where 
\be
\eta_i^t(s) = \sqrt{nP_\ell}    \frac{\exp\left({s_i\frac{\sqrt{nP_\ell}}{\tau_t^2}}\right)}   {\sum_{j\in\sec(i)}\exp\left({s_j\frac{\sqrt{nP_\ell}}{\tau_t^2}}\right)}, \quad 1 \leq i \leq ML.
 \label{eq:eta_update}
\ee
Here the notation $j\in\sec(i)$ refers to all indices $j$ in the same section as $i$. (Note that there are $M$ indices in each section.)  At the end of each step $t$, $\beta_i^{t}/\sqrt{nP_\ell}$ may be interpreted as the updated  posterior probability of the $i$th entry being the non-zero one in its section.

The constants $\tau_t^2$  are specified by the following scalar recursion called ``state evolution" (SE):
\begin{align}
\tau^2_{0}  =\sigma^2+P,  \qquad  \tau^2_{t}  =  \sigma^2 + P(1-x(\tau_{t-1})),  \quad  t \geq 1,
\label{eq:taut_def}
\end{align}
where 
\be
\begin{split}
 x(\tau) := \sum_{\ell=1}^{L} \frac{P_\ell}{P} \, \expec \left[
\frac{e^{ \frac{\sqrt{n P_\ell}}{\tau} \, (U^{\ell}_1  + \frac{\sqrt{n P_\ell}}{\tau})} }{e^{ \frac{\sqrt{n P_\ell}}{\tau} \, (U^{\ell}_1  + \frac{\sqrt{n P_\ell}}{\tau})}  + \sum_{j=2}^M 
e^{ \frac{\sqrt{n P_\ell}}{\tau}U^{\ell}_j }} \right].
\end{split}
\label{eq:xt_tau_def}
\ee
In \eqref{eq:xt_tau_def}, $\{ U^\ell_j\}$ are i.i.d.\ $\mc{N}(0,1)$ random variables for $j\in [M], \ \ell \in [L]$. The significance of the SE  parameters $\tau_t^2$ is discussed in Section \ref{sec:pa}.  In Section \ref{sec:tau}, we use an online approach to accurately compute the $\tau_t^2$ values rather than pre-computing them via \eqref{eq:xt_tau_def}.

At the end of $T$ iterations, the decoded message vector $\widehat{\beta}$ is produced by setting the maximum value in section $\ell$ of $\beta^T$ to $\sqrt{nP_\ell}$ and the remaining entries to zero, for $1 \leq \ell \leq L$.

\emph{Error rate of the AMP decoder}:  We measure the \emph{section error rate} $\Es$ as
\be
    \Es = \frac{1}{L}\sum_{\ell=1}^{L}
        \mathbf{1}{\left\{ \widehat{\beta}_\ell \neq \beta_{\ell} \right\}}
\ee
Assuming a uniform mapping between the input bitstream and the non-zero locations in each section, each section error will cause approximately half of
the bits it represents to be incorrect, leading to a \emph{bit error rate} $\Eb \approx \frac{1}{2} \Es$. 

Another figure of merit is the \emph{codeword error rate} $\Ec$, which
estimates the probability $\mathbb{P}(\widehat{\beta} \neq \beta)$.  If the
SPARC is used  to transmit a large number of messages (each via a length $n$
codeword), $\Ec$ measures the fraction of codewords that are decoded with one
or more section errors. The codeword error rate is insensitive to where and how
many section errors occur within a codeword when it is decoded incorrectly.

At finite code lengths, the choice of a good power allocation crucially depends
on whether we want to minimize $\Es$ or $\Ec$. As we will  see in the next
section,  a power allocation that yields reliably low section error rates  may
result in a high codeword error rate, and vice versa. In this paper, we will
mostly focus on obtaining the best possible section error rate, since in
practical applications a high-rate outer code could readily correct a small
fraction of section errors to give excellent codeword error rates as well.
Further, the bit error rate (which is approximately half the section error
rate) is useful to compare with  other channel coding approaches,  where
it is a common figure of merit.


\subsection{Organization of the paper and main contributions}

In the rest of the paper, we describe several techniques to improve the finite length error performance and reduce the complexity of AMP decoding. The  sections are organized as follows. 
\begin{itemize}
    \item In Section~\ref{sec:pa}, we introduce an iterative power allocation algorithm  that gives improved error performance with fewer tuning parameters than other power allocation schemes.
        
     \item In Section~\ref{sec:error-concentration}, we analyze the effects of the code parameters $L, M$ and the power allocation on error performance and its concentration around the value predicted by state evolution.
     
    \item In Section~\ref{sec:tau}, we describe how an online estimate of the key SE parameter $\tau_t^2$   improves error performance and  allows a new early-stopping criterion. Furthermore, the online estimate enables us to accurately estimate the actual section  error rate at the end of the decoding process. 
        
    \item In Section~\ref{sec:fsk}, we derive simple expressions to  estimate $\Es$ and $\Ec$ given the rate and power allocation.
        
    \item In Section~\ref{sec:ldpc} we compare the error performance of AMP-decoded SPARCs  to LDPC-based coded modulation schemes used in the WiMAX standard.
    
    \item In Section \ref{sec:ldpc-outer}, we describe how partial outer codes can be used in conjunction with AMP decoding.  We propose a three-stage decoder consisting of AMP decoding, followed by outer code decoding, and finally, AMP decoding once again.
     We show that by covering only a fraction of sections of the message $\beta$ with an outer code,  the three-stage decoder can correct errors even in the sections not covered by the outer code. This results in bit-error curves with a steep waterfall behavior.
\end{itemize}

The  main technical contributions of the paper are the iterative power allocation algorithm (Section~\ref{sec:pa}) and the  three-stage decoder with an outer code (Section \ref{sec:ldpc-outer}).  The other sections describe how various choices of code parameters influence the finite length error performance,  depending on whether the objective is to minimize the section error rate or the codeword error rate. We remark that the focus in this paper is on improving the finite length performance using the standard SPARC  construction with power allocation. Optimizing the finite length performance of spatially-coupled SPARCs considered in \cite{BarbKrz15,BarbSK2015}  is an interesting research direction, but one that is beyond the scope of this paper. 
\section{Power Allocation} \label{sec:pa}

Before introducing the power allocation scheme, we briefly give some intuition about the AMP update rules \eqref{eq:z_update}--\eqref{eq:eta_update}, and the SE recursion in \eqref{eq:taut_def}--\eqref{eq:xt_tau_def}. The update step \eqref{eq:beta_update} to generate each estimate of $\beta$ is underpinned by the following key property: after step $t$, the ``effective observation" $\beta^t + A^* z^t$ is approximately distributed as $\beta + \tau_t Z$, where $Z$ is standard normal random vector independent of $\beta$.  Thus $\tau_t^2$ is the effective noise variance at the end of step $t$. Assuming that the above distributional property holds,  $\beta^{t+1}$ is  just the Bayes-optimal estimate of $\beta$ based on the effective observation. The entry
$\beta^{t+1}_i$ is proportional to the posterior probability of the $i$th entry
being the non-zero entry in its section.
 
We see from \eqref{eq:taut_def} that the effective noise variance $\tau_{t}^2$ is the sum of two terms. The first is the channel noise variance $\sigma^2$.  The other term $P(1-x(\tau_{t-1}))$ can be interpreted as the interference due to the undecoded sections in $\beta^t$. Equivalently, $x(\tau_{t-1})$ is the expected power-weighted fraction of sections which are correctly decodable at the end of step $t$.

The starting point for our power allocation design is the following result from \cite{Rush2017},
which gives analytic upper and lower bounds for $x(\tau)$ of \eqref{eq:taut_def}.
\begin{lem}
\cite[Lemma 1(b)]{Rush2017b}
Let $\nu_\ell := \frac{ L P_\ell}{R \tau^2 \ln 2 }$.  For sufficiently large $M$, and for any $\delta \in (0,1)$,
\begin{align}
x(\tau) & \leq \sum_{\ell=1}^L \frac{P_\ell}{P} \left[ \mathbf{1}\left\{ \nu_\ell > 2 - \delta  \right\} + M^{-\kappa_1 \delta^2}  \mathbf{1}\left\{ \nu_\ell \leq 2 -\delta  \right\} \right],   \label{eq:xub_asym}  \\
x(\tau) & \geq  \left(1- \frac{M^{-\kappa_2 \delta^2}}{\delta \sqrt{\ln M}} \right) \sum_{\ell=1}^{L} \frac{P_\ell}{P} \,
\mathbf{1}\left\{ \nu_\ell > 2 + \delta \right\}.  \label{eq:xlb_asym} 
\end{align}
where $\kappa_1, \kappa_2$ are universal positive constants.
\label{lem:xlem}
\end{lem}
As the constants $\kappa_1, \kappa_2$ in  \eqref{eq:xub_asym}--\eqref{eq:xlb_asym} are not precisely specified, for designing power allocation  schemes, we use the following approximation for $x(\tau)$:
\be
x(\tau) \approx \,   \sum_{\ell=1}^{L} \frac{P_\ell}{P} \,
\mathbf{1}\left\{ LP_\ell >  2R \tau^2 \ln 2  \right\}.
\label{eq:lemma1b}
\ee

This approximate  version, which is increasingly accurate as  $L,M$ grow large, is useful for gaining intuition about suitable power allocations. Indeed,  if the effective noise variance after step $t$ is $\tau_t^2$, then  \eqref{eq:lemma1b} says that any section $\ell$ whose normalized power $L P_\ell$ is larger than the threshold  $2 R \tau^2_t \ln 2$ is likely to be decodable correctly in step $(t+1)$, i.e., in  $\beta^{t+1}$, the probability mass within the section will be concentrated on the correct non-zero entry.   For a given power allocation,  we can iteratively estimate the SE parameters $(\tau_t^2, x(\tau_t^2))$ for each $t$ using  the lower bound in \eqref{eq:lemma1b}. This provides a way to quickly check whether or not a given power allocation will lead to reliable decoding in the large system limit. For reliable decoding at a given rate $R$,  the effective noise variance  given by  $\tau_t^2= \sigma^2+ P(1- x(\tau_{t-1}))$ should decrease with $t$ until it reaches a value close to $\sigma^2$ in a finite number of iterations.  Equivalently, $x(\tau_{t})$ in \eqref{eq:xt_tau_def} should  increase to a value very close to $1$.

For a  rate $R < \mc{C}$, there are infinitely many power allocations for which \eqref{eq:lemma1b} predicts successful decoding in the large system limit. However, as illustrated below, their finite length error performance may differ significantly. Thus the  key question addressed in this section is: \emph{how do we choose a power allocation that gives the lowest section error rate}?

The exponentially-decaying power allocation given by 
 \be
P_\ell = \frac{P(2^{2\C/L}-1)}{1-2^{-2\C}} 2^{-2\C\ell/L}, \quad \ell \in [L], \label{eq:exp_decayPA}
\ee 
was proven in  \cite{Rush2017}  to be capacity-achieving in the large system limit, i.e., it was shown that  the section error rate $\Es$ of the AMP decoder converges almost surely to $0$ as $n \to \infty$, for any $R < \mc{C}$. However, it does not perform well at practical block lengths, which motivated the search for alternatives. We  now evaluate it in the context of \eqref{eq:lemma1b} to better explain the development of a new power allocation scheme.

\begin{figure}[t]
    \centering
    \includegraphics[width=0.7\columnwidth]{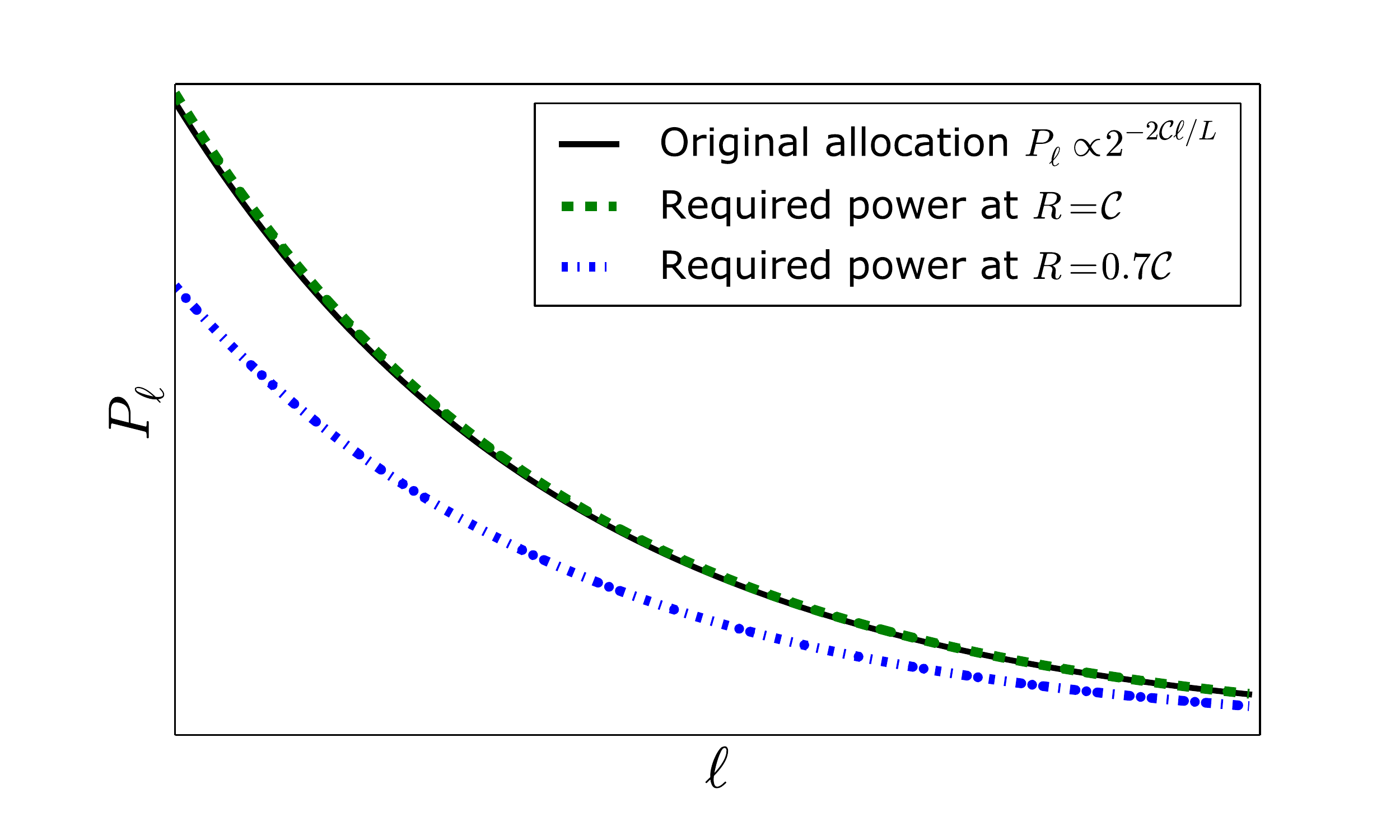}
\caption{\small The dashed lines show the minimum required power in section for successful decoding when $R=\mc{C}$  (above), and $R=0.7 \mc{C}$ (below), where $\mc{C} =2$ bits.  The solid line shows the exponentially-decaying power allocation in  \eqref{eq:exp_decayPA}.}
\label{fig:pa_origpa}
\vspace{-8pt}
\end{figure}

Given a power allocation, using \eqref{eq:lemma1b} one can compute the minimum required power for any section $\ell \in [L]$ to decode, assuming that the sections with higher power have decoded correctly. 
The dashed lines in Figure \ref{fig:pa_origpa} shows the minimum power  required for each section to decode (assuming the exponential allocation of \eqref{eq:exp_decayPA}  for the previous sections), for $R=\mc{C}$ and $R=0.7 \mc{C}$. The figure shows that the power allocation in \eqref{eq:exp_decayPA} matches (up to order $\frac{1}{L}$ terms) with the minimum required power when $R=\mc{C}$. However, for $R=0.7 \mc{C}$, we see that the exponentially-decaying allocation allocates significantly more power to the earlier sections than the minimum required, compared to later sections. This leads to relatively high section error rates, as shown in Figure \ref{fig:pa_perf_comparison}.

Figure \ref{fig:pa_origpa} shows that the total power  allocated by the minimal
power allocation at  $R=0.7 \mc{C}$ is significantly less than the available
power $P$.  Therefore, the key question is: how do we balance the allocation of
available power between the various sections to minimize the section error rate? 
Allocating excessive power to the earlier sections ensures
they decode reliably early on, but then there will not be sufficient power left
to ensure reliable decoding in the final sections. This is the reason for the
poor finite length performance of the exponentially-decaying allocation.
Conversely, if the power is spread too evenly then no section particularly
stands out against the noise, so it is hard for the decoding to get started,
and early errors can cause cascading failures as subsequent sections are also
decoded in error.  

\begin{figure}[t]
    \centering
    \includegraphics[width=0.7\columnwidth]{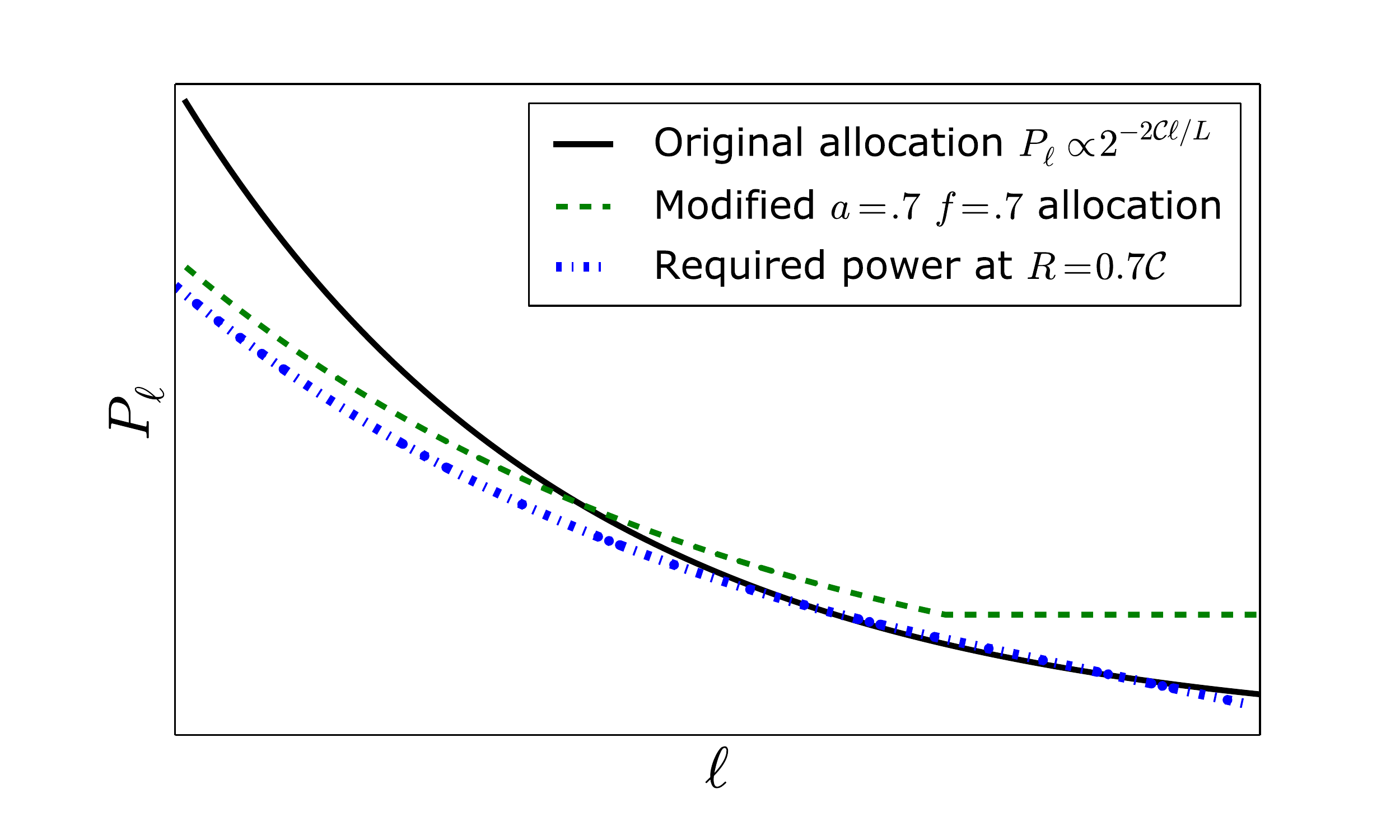}
    \caption{\small The modified power allocation with $a=f=0.7$ results in slightly more than the minimum power required for the first $70 \%$ of sections; the remaining available power is allocated equally among the last $30\%$ of sections. The original allocation with $P_\ell \propto 2^{-2 \mc{C} \ell /L}$ is also shown for comparison. }     \label{fig:pa_af}
    \vspace{-8pt}
\end{figure}

This trade-off motivated the following modified exponential
power allocation proposed in \cite{Rush2017}:
\be P_\ell = \begin{cases} \kappa \cdot 2^{-2a\C \ell/L} & 1 \leq \ell \leq fL,\\ 
\kappa \cdot 2^{-2a\C f}   & fL+1 \leq \ell
\leq L,  \end{cases} \label{eq:PA_af} \ee 
where the normalizing constant $\kappa$ is chosen to ensure that $\sum_{\ell=1}^{L} P_\ell = P$.
In \eqref{eq:PA_af}, the parameter $a$ controls the steepness of the  exponential allocation, while the parameter $f$ flattens the allocation after the first fraction $f$ of the sections. Smaller choices of $a$ lead to less power allocated to
the initial sections, making a larger amount available to the later sections. Similarly, smaller values of $f$ lead to more power  allocated to the final sections. See Figure~\ref{fig:pa_af} for an illustration.

While this allocation improves the section error rate by a few orders of
magnitude (see \cite[Fig. 4]{Rush2017}), it requires costly
numerical optimization of $a$ and $f$. A good starting point is to use
$a=f=R/\C$, but further optimization is generally necessary. This motivates
the need for a fast power allocation algorithm with fewer tuning parameters.

\subsection{Iterative power allocation}
\label{sec:pa:iterative}
We now describe a simple parameter-free iterative algorithm to design a power allocation.   The $L$ sections of the SPARC  are divided into $B$ \emph{blocks} of $L/B$ sections each. Each section within a block is allocated the same power.  For example, with $L=512$ and $B=32$, there are $32$ blocks  with $16$ sections per block. The algorithm sequentially allocates power to each of the $B$ blocks as follows. Allocate the minimum power to the first block of sections so that they can be decoded in the first iteration when $\tau_0^2=\sigma^2+P$. Using \eqref{eq:lemma1b}, we set the  power in each section of the first block to
\[ P_\ell = \frac{2 R\tau_0^2 \ln2}{L}, \quad 1\le\ell\le\frac{L}{B}.\]  Using \eqref{eq:lemma1b} and \eqref{eq:taut_def}, we then estimate
$\tau_1^2=\sigma^2+(P -  B P_{1})$. Using this value, allocate the minimum
required power for the second block of sections to decode, i.e.,
$P_\ell=2\ln2R\tau_1^2/L$ for $\frac{L}{B}+1\le\ell \le \frac{2L}{B}$.
If we sequentially allocate power in this manner to  each of the $B$ blocks, then the total power allocated by this scheme will be strictly less than $P$ whenever $R < \mc{C}$.  We  therefore modify the scheme as follows.  

\begin{figure}[t]
    \centering
    \includegraphics[width=0.3\columnwidth]{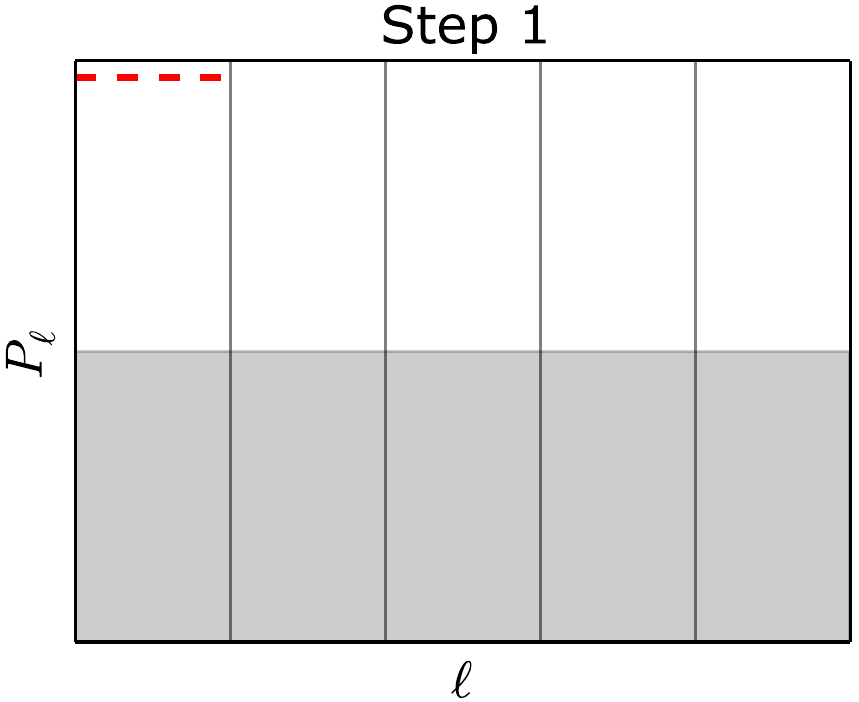}
    \includegraphics[width=0.3\columnwidth]{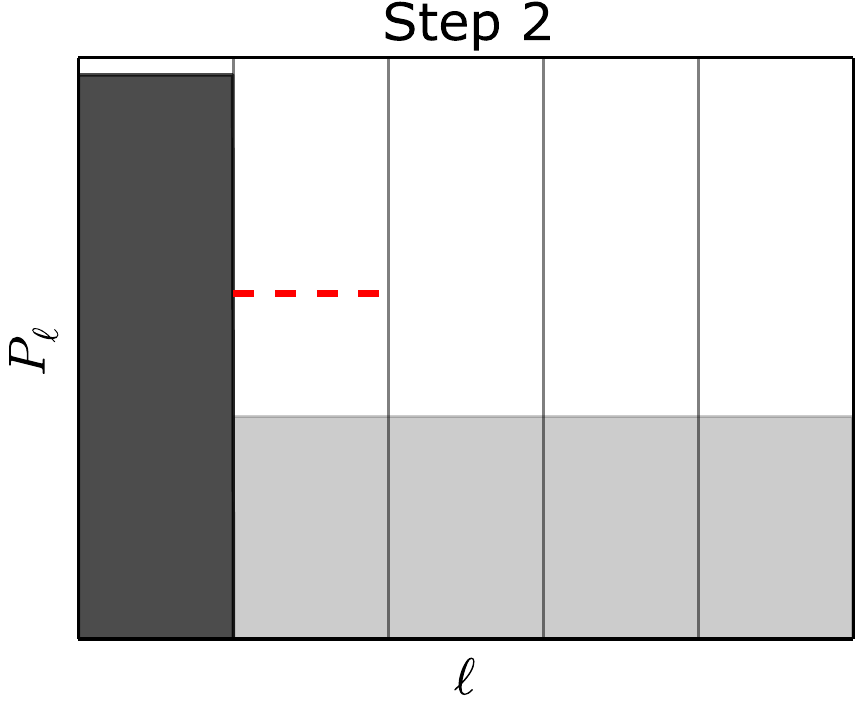}
    \includegraphics[width=0.3\columnwidth]{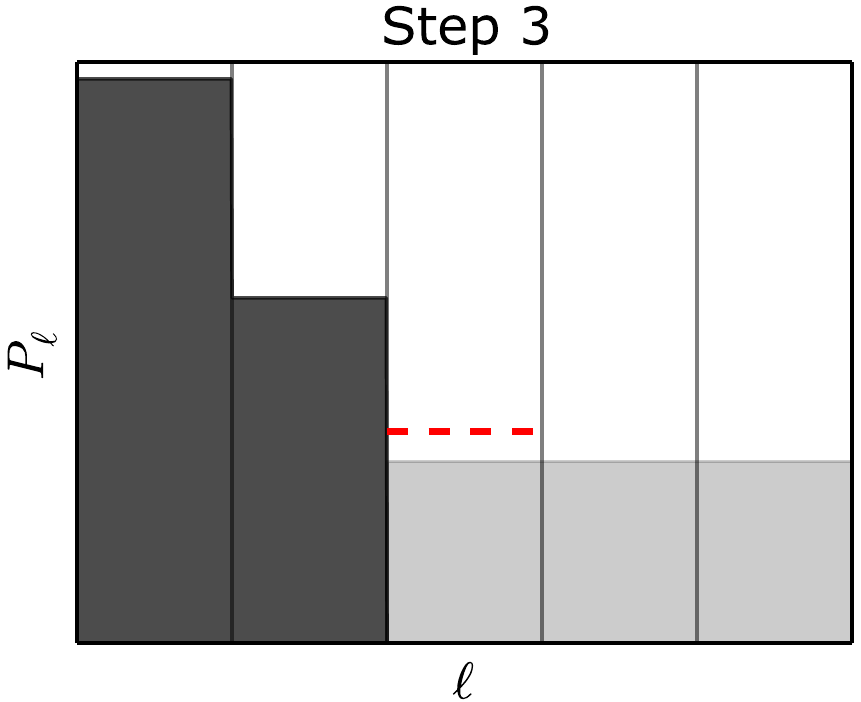}
    \includegraphics[width=0.3\columnwidth]{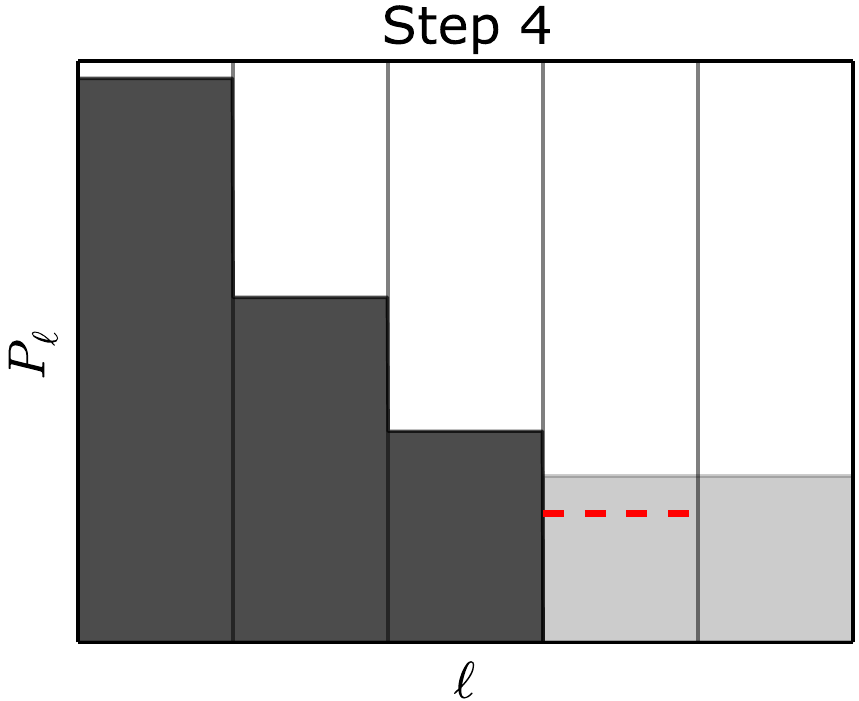}
    \includegraphics[width=0.3\columnwidth]{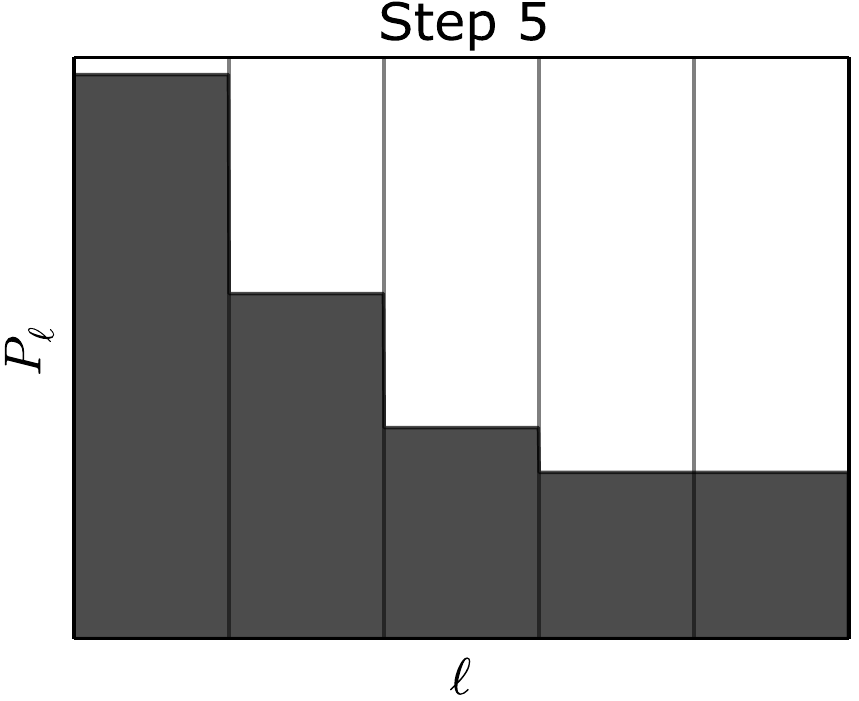}
    \caption{\small Example illustrating the iterative power allocation algorithm with $B=5$. In each step, 
        the height of the light gray region represents the allocation that distributes the remaining power equally
        over all the remaining sections. The dashed red line indicates the minimum power required for decoding the current block of sections.
       The dark gray bars represent the power that has been allocated at the beginning of the current step. }
    \label{fig:pa_lemma1_animation}
\end{figure}

\begin{algorithm}[t]
\caption{Iterative power allocation routine}%
\label{alg:iterative-pa}
\begin{algorithmic}
    \REQUIRE $L$, $B$, $\sigma^2$, $P$, $R$ such that $B$ divides $L$.
    \STATE Initialise $k \leftarrow  \frac{L}{B}$
    \FOR{$b=0$ to $B-1$}
        \STATE $P_{\text{remain}} \leftarrow P - \sum_{\ell=1}^{bk}P_\ell $
        \STATE $\tau^2 \leftarrow \sigma^2 + P_{\text{remain}}$
        \STATE $P_{\text{block}} \leftarrow 2 \ln(2) R \tau^2 / L$
        \IF{$P_{\text{remain}}/(L-bk) > P_{\text{block}}$}
            \STATE $P_{bk+1},\ldots,P_L \leftarrow P_{\text{remain}}/(L-bk)$
            \BREAK
        \ELSE
        \STATE $P_{bk+1},\ldots,P_{(b+1)k} \leftarrow P_{\text{block}}$
        \ENDIF
    \ENDFOR
    \RETURN $P_1,\ldots,P_L$
\end{algorithmic}
\end{algorithm}

\begin{figure}[t]
    \centering
    \includegraphics[width=0.8\columnwidth]{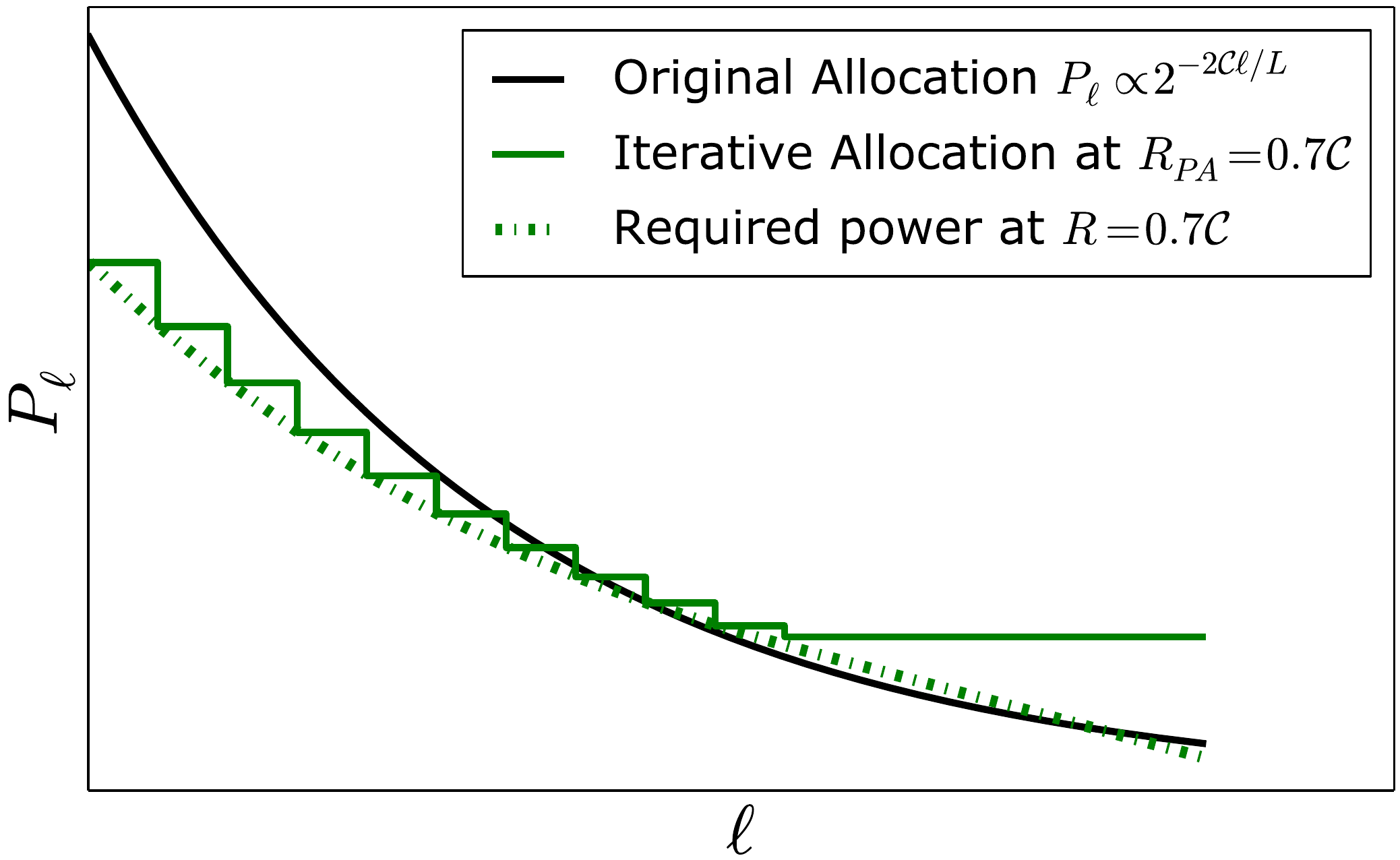}
    \caption{\small Iterative allocation, with $L=512$, and $B=16$ blocks. Flattening occurs at the
    11th block.}%
    \label{fig:pa_lemma1_shape}
\end{figure}

For $1 \leq b \leq B$, to allocate power to the $b$th block of sections assuming that the first $(b-1)$ blocks have been allocated, we compare the two options and choose the one that allocates higher power to the block: i) allocating the minimum required power (computed as above) for the $b$th block of sections to decode; ii) allocating the remaining available power equally to sections in blocks $b, \ldots, B$, and terminating the algorithm. This gives a flattening in the final
blocks similar to the allocation in \eqref{eq:PA_af}, but without requiring a specific parameter that determines where the flattening begins.  The iterative power allocation routine is described in Algorithm~\ref{alg:iterative-pa}. Figure~\ref{fig:pa_lemma1_animation} shows a toy example building up the power allocation for $B=5$, where flattening is seen to occur in step 4.
Figure~\ref{fig:pa_lemma1_shape} shows a more realistic example with $L=512$ and $R=0.7 \mc{C}$.
\begin{figure}[t]
    \centering
    \includegraphics[width=0.95\columnwidth]{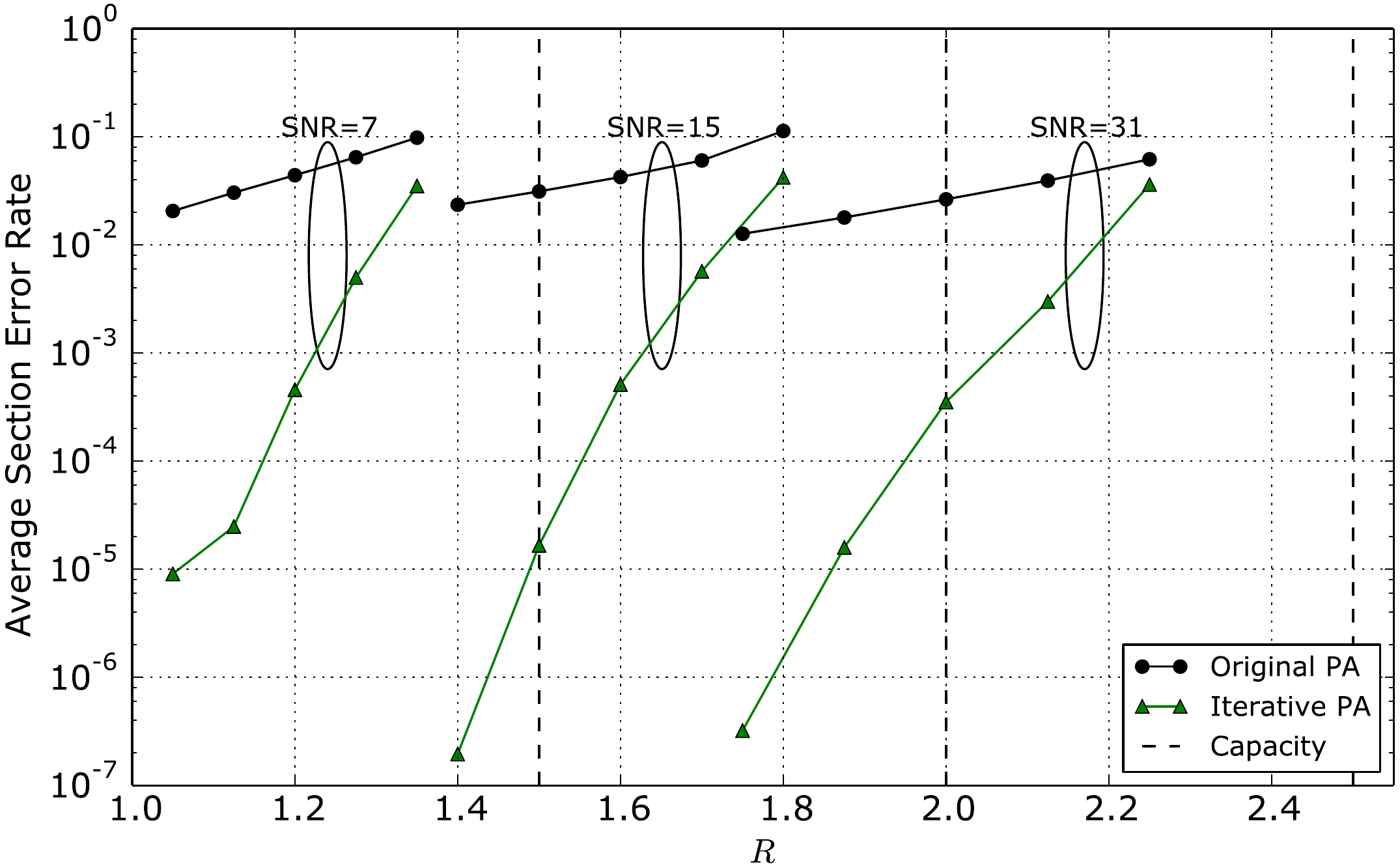}
    \caption{\small AMP section error rate $\Es$ vs $R$ at $\snr=7,15,31$,
    corresponding to $\mc{C}=1.5,2,2.5$ bits (shown with dashed vertical lines).
    At each $\snr$, the section error rate is reported for rates $R/\mc{C}=0.70,0.75,0.80,0.85,0.90$.
    The SPARC parameters are $M=512,L=1024$. The top black curve shows the
    $\Es$ with $P_\ell \propto 2^{-2\mc{C}\ell/L}$. The lower green curve
    shows $\Es$ for the iterative power allocation, with $B=L$ and $R_\text{PA}$
    numerically optimized. (See Sec. \ref{subsec:conc-pa} for a discussion of $R_{\text{PA}}$.)}
    \label{fig:pa_perf_comparison}
    \vspace{-8pt}
\end{figure}

\emph{Choosing $B$}: By construction, the iterative power allocation scheme specifies the number of iterations of the AMP decoder in the large system limit.  This is given by the number of blocks with distinct powers; in particular the number of iterations (in the large system limit) is of the order of  $B$.  For finite code lengths, we find that it is better to use a termination criterion for the decoder based on the estimates generated by the algorithm. This criterion is described in Sec. 
\ref{sec:tau}. This data-driven termination criterion allows us to choose the number of blocks $B$ to be as large as $L$. We found that choosing $B=L$, together with the termination criterion in Sec. \ref{sec:tau}, consistently gives a small improvement in error performance (compared to other choices of $B$), with no additional time
or memory cost.  

Additionally, with $B=L$, it is possible to quickly determine a pair
$(a,f)$ for the modified exponential allocation in \eqref{eq:PA_af} which gives a nearly identical
allocation to the iterative algorithm. This is done by first setting $f$ to obtain the same
flattening point found in the iterative allocation, and then searching for an
$a$ which matches the first allocation coefficient $P_1$ between the iterative
and the modified exponential allocations. Consequently, any simulation results
obtained for the iterative power allocation could also be obtained using a
suitable $(a,f)$ with the modified exponential allocation, without having to
first perform a costly numerical optimization over $(a,f)$. 

Figure~\ref{fig:pa_perf_comparison} compares the error performance of the exponential and iterative power allocation schemes discussed above for different values of $R$ at $\snr=7,15,31$. The iterative power allocation yields significantly improved $\Es$ for rates away from capacity when compared to the original exponential allocation, and additionally outperforms the modified exponential allocation results reported in \cite{Rush2017}.

For the experiments in Figure~\ref{fig:pa_perf_comparison}, the value for $R$ used in constructing the iterative allocation (denoted by $R_{PA}$) was optimized numerically. Constructing an iterative allocation with  $R=R_{PA}$ yields good results, but due to finite length concentration effects,  the $R_{PA}$  yielding the smallest average error rate may be slightly different from the communication rate $R$. The effect of $R_{PA}$ on the  concentration of error rates  is discussed in Section~\ref{subsec:conc-pa}. We emphasize that this optimization over $R_{PA}$ is simpler than numerically optimizing the pair $(a,f)$ for the modified exponential allocation. Furthermore, guidelines for choosing $R_{PA}$ as a function of $R$ are given in Section~\ref{subsec:conc-pa}.

\section{Error Concentration Trade-offs}
\label{sec:error-concentration}

In this section, we discuss how the choice of  SPARC design parameters can influence the trade-off between the `typical'  value of section error rate  and concentration of actual error rates around the typical values. The typical section error rate refers to that predicted by state evolution (SE).  Indeed,  running the SE equations
\eqref{eq:taut_def}--\eqref{eq:xt_tau_def}  until convergence gives the following prediction for the section
error rate:
\be
\mc{E}^{\text{SE}}_{\sec} := 1 - \frac{1}{L}\sum_{\ell= 1}^{L} \expec \left[
    \frac{e^{
                \frac{\sqrt{nP_\ell}}{\tau_T}
                \left(U_1^\ell + \frac{\sqrt{nP_\ell}}{\tau_T}\right)
            }
         }
         {e^{
                \frac{\sqrt{nP_\ell}}{\tau_T}
                \left(U_1^\ell + \frac{\sqrt{nP_\ell}}{\tau_T}\right)
            }
          +
          \sum_{j=2}^M e^{\frac{\sqrt{nP_\ell}}{\tau_T}U_j^\ell}
         }
\right],
\label{eq:SE_pred}
\ee
where $\tau_T^2$ denotes the value in the final iteration.
The concentration refers to how close the SE prediction $\mc{E}^{\text{SE}}_{\sec}$ is to the observed section error rate.

As we describe below,  the choice of SPARC parameters $(L,M)$  and the power allocation both determine a trade-off between obtaining 
a low value for $\mc{E}^{\text{SE}}_{\sec}$,  and concentration of the actual section error rate around  $\mc{E}^{\text{SE}}_{\sec}$.  This trade-off is of particular interest when applying an outer code to the SPARC,
as considered in Section~\ref{sec:ldpc-outer}, which may be able to reliably handle only a small number of section errors.

\subsection{Effect of $L$ and $M$ on concentration} \label{sec:lvsm}

Recall  from \eqref{eq:Rdef} that the code length $n$ at a given rate $R$
is determined by the choice of $L$ and $M$ according to the relationship
$nR=L\log M$. In general, $L$ and $M$ may be chosen freely to meet a desired
rate and code length. 

\begin{figure}[t]
    \centering
    \includegraphics[width=0.95\columnwidth]{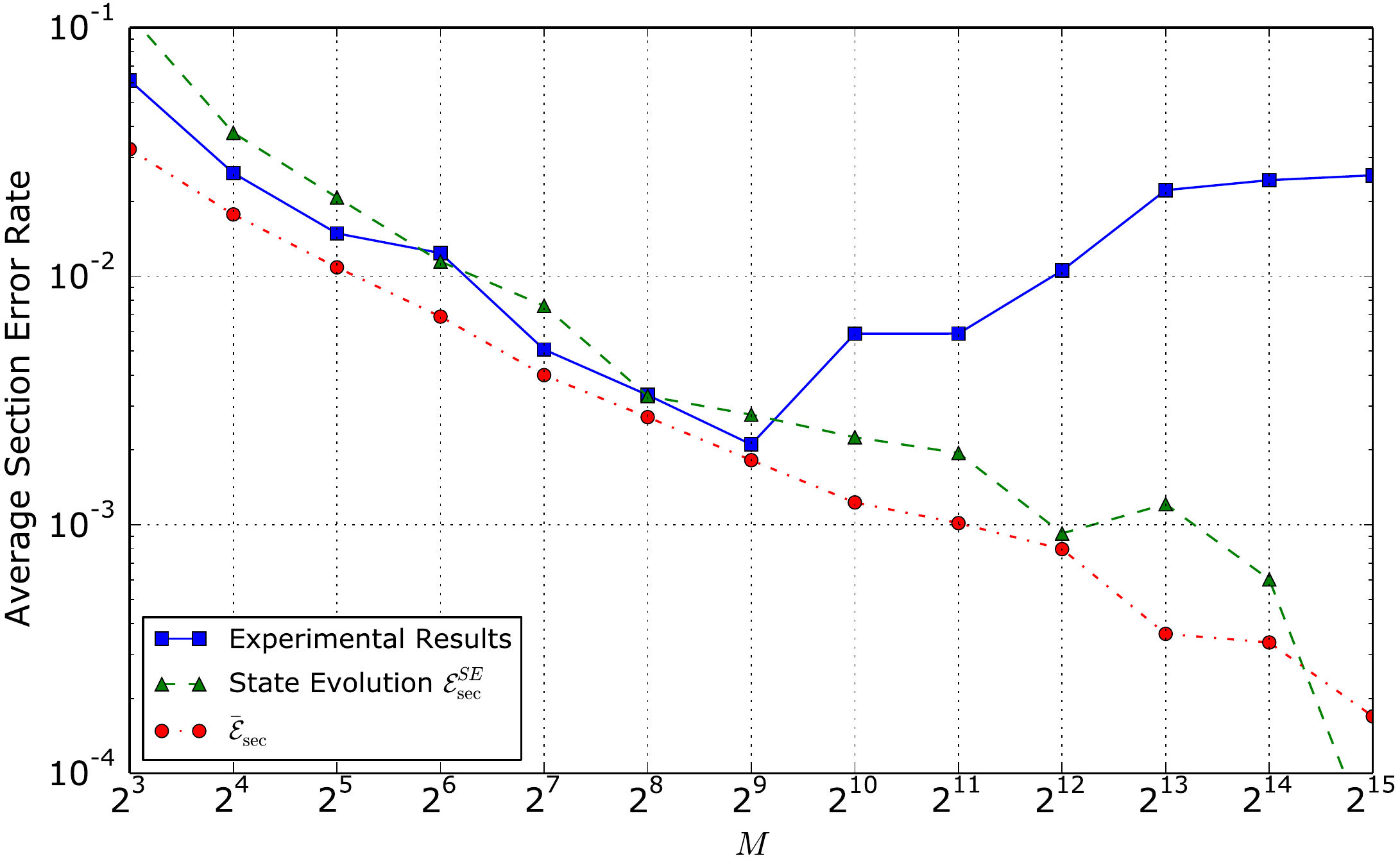}
    \caption{\small AMP error performance with increasing $M$, for
        $L=1024$, $R=1.5$, and $\frac{E_b}{N_0}=5.7$ dB (2 dB from Shannon limit). See Section~\ref{sec:fsk}
        for details of $\bar{\mc{E}}_{\text{sec}}$.%
    }
\label{fig:lvsm_ser}
\vspace{-8pt}
\end{figure}

To understand the effect of  increasing $M$, consider Figure \ref{fig:lvsm_ser} which shows the error performance of a SPARC with $R=1.5, L=1024$, as we increase the value of $M$. From \eqref{eq:Rdef}, the code length
$n$ increases logarithmically with $M$. We observe that the  section error rate (averaged over $200$ trials)
decreases with $M$ up to $M=2^9$, and then starts increasing. This is in sharp
contrast to the SE prediction \eqref{eq:SE_pred} (plotted using a dashed line in Figure \ref{fig:lvsm_ser}) which keeps decreasing as   $M$ is increased.  

This divergence between the actual section error rate and the SE prediction for large $M$  is
due to large fluctuations  in the number of section errors across trials.
Recent work on the error exponent of SPARCs with AMP decoding
shows that the concentration of error rates near the SE prediction is strongly
dependent on both $L$ and $M$. For $R < \mc{C}$, \cite[Theorem 1]{Rush2017b}
shows that for any $\e >0$, the section error rate $\Es$ satisfies
\be
\begin{split}
    &\mathbb{P} \left( \mc{E}_{\sec}  > \mc{E}^{\text{SE}}_{\sec} +  \e \right) \leq 
    K_{T} \, e^{\frac{-\kappa_{T} L}{(\log M)^{2T-1}}
    \left(\frac{\e  \ln(1 + \snr)}{4 (1 + \snr)} -  f(M) \right)^2},
 \label{eq:pezero}
\end{split}
\ee
where $T$ is the number of iterations until state evolution convergence,
$\kappa_T, K_T$ are constants depending on $T$, and $f(M)=
\frac{M^{-\kappa_2 \delta^2}}{\delta \sqrt{\ln M}}$ is a quantity that tends to zero
with growing $M$.  For any power allocation, $T$ increases as $R$ approaches
$C$. For example,  $T \propto 1/\log (\mc{C}/R)$ for the exponential power allocation.
We observe that the deviation probability bound on the RHS of \eqref{eq:pezero}
depends on the ratio $L / (\log M)^{2T-1}$.

\begin{figure}[t]
    \centering
    \includegraphics[width=0.95\columnwidth]{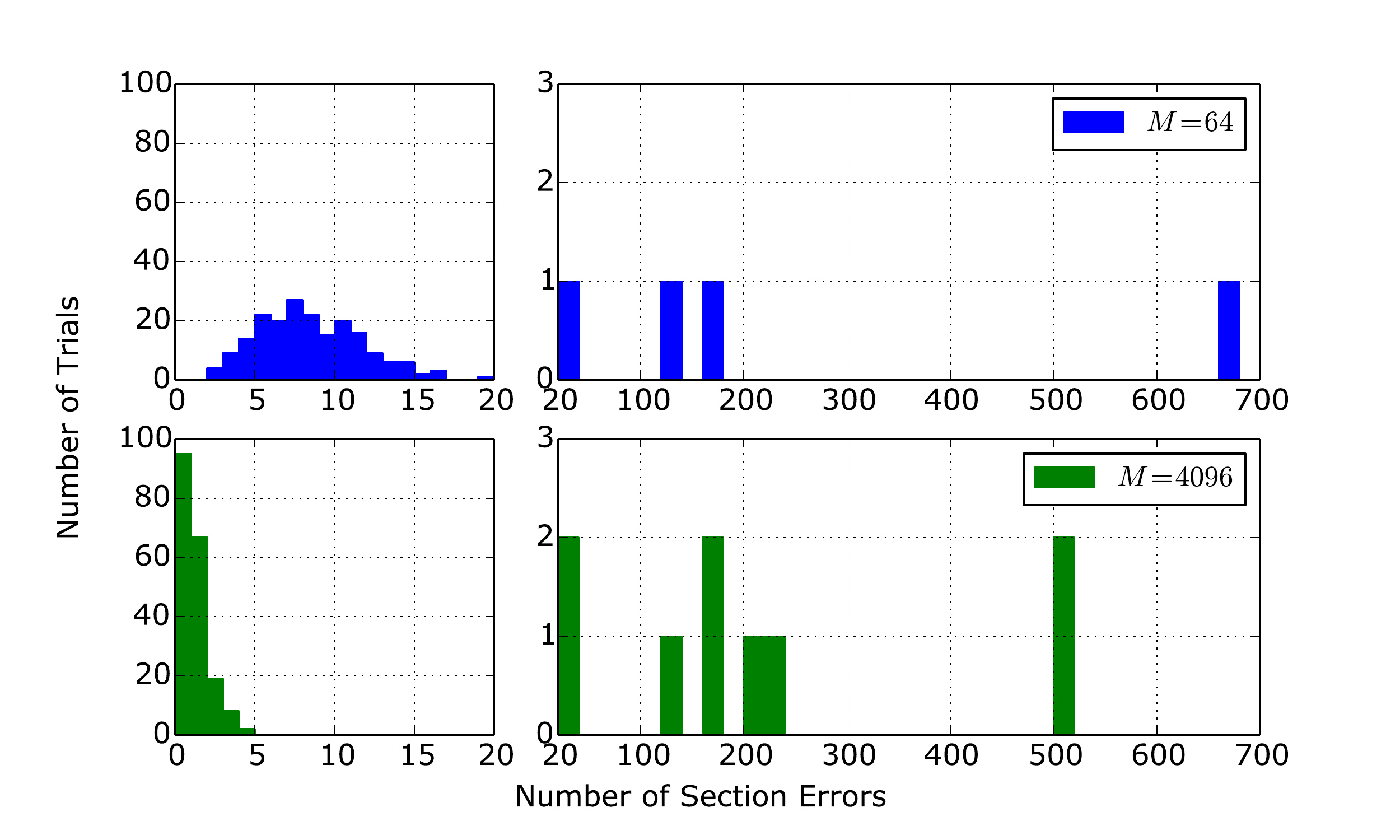}
    \caption{\small Histogram of AMP section errors over $200$ trials $M=64$ (top) and $M=4096$
        (bottom), with $L=1024$, $R=1.5$, $\frac{E_b}{N_0}=5.7$dB. The left panels highlight
        distribution of errors around low section error counts, while the right
        panels show the distribution around high-error-count events. As shown
        in Figure~\ref{fig:lvsm_ser}, both cases have an average section error
        rate of around $10^{-2}$.}%
\label{fig:lvsm_hist}
\end{figure}

In our experiments, $T$  is generally on the order of a few tens. Therefore,
keeping $L$ constant,  the probability of large deviations from the SE
prediction $\mc{E}^{\text{SE}}_{\sec}$ increases with $M$.   This leads to the
situation shown in Figure~\ref{fig:lvsm_ser}, which shows that the SE
prediction  $\mc{E}^{\text{SE}}_{\sec}$ continues to decrease with $M$, but
beyond a certain value of $M$, the observed average section error rate becomes
progressively worse due to loss of concentration. This is caused by a small
number of trials with a very large number of section errors, even as the
majority of trials experience lower and lower error rates as $M$ is increased.
This effect can be clearly seen in  Figure~\ref{fig:lvsm_hist}, which 
compares the histogram of section error rates over $200$ trials for $M=64$ and
$M=4096$. The distribution of errors is clearly different, but both cases have
the same average section error rate due to the poorer concentration for
$M=4096$.


To summarize,  given $R, \snr$, and $L$, there is an optimal $M$ that minimizes
the empirical section error rate. Beyond this value of $M$, the benefit from
any further increase is outweighed by the loss of concentration. For a given
$R$, values of $M$ close to $L$ are a good starting point for optimizing the
empirical section error rate, but obtaining closed-form estimates of the
optimal $M$  for a given $L$ is still an open question.

For  fixed $L, R$, the optimal value of $M$  increases with $\snr$. This effect
can be seen in the results of Figure~\ref{fig:ldpc-comparison-r10}, where there
is an inversion in the order of best-performing $M$ values as $E_b/N_0$
increases. This is because as $\snr$ increases, the number of iterations $T$
for SE to converge decreases. A smaller $T$ mitigates the effect of larger $M$
in the large deviations bound of \eqref{eq:pezero}. In other words, a larger
$\snr$ leads to better error rate concentration around the SE prediction, so
larger values of $M$ are permissible before the performance starts degrading. 

\subsection{Effect of power allocation on concentration} \label{subsec:conc-pa}

\begin{figure}[t]
    \centering
    \includegraphics[width=0.95\columnwidth]{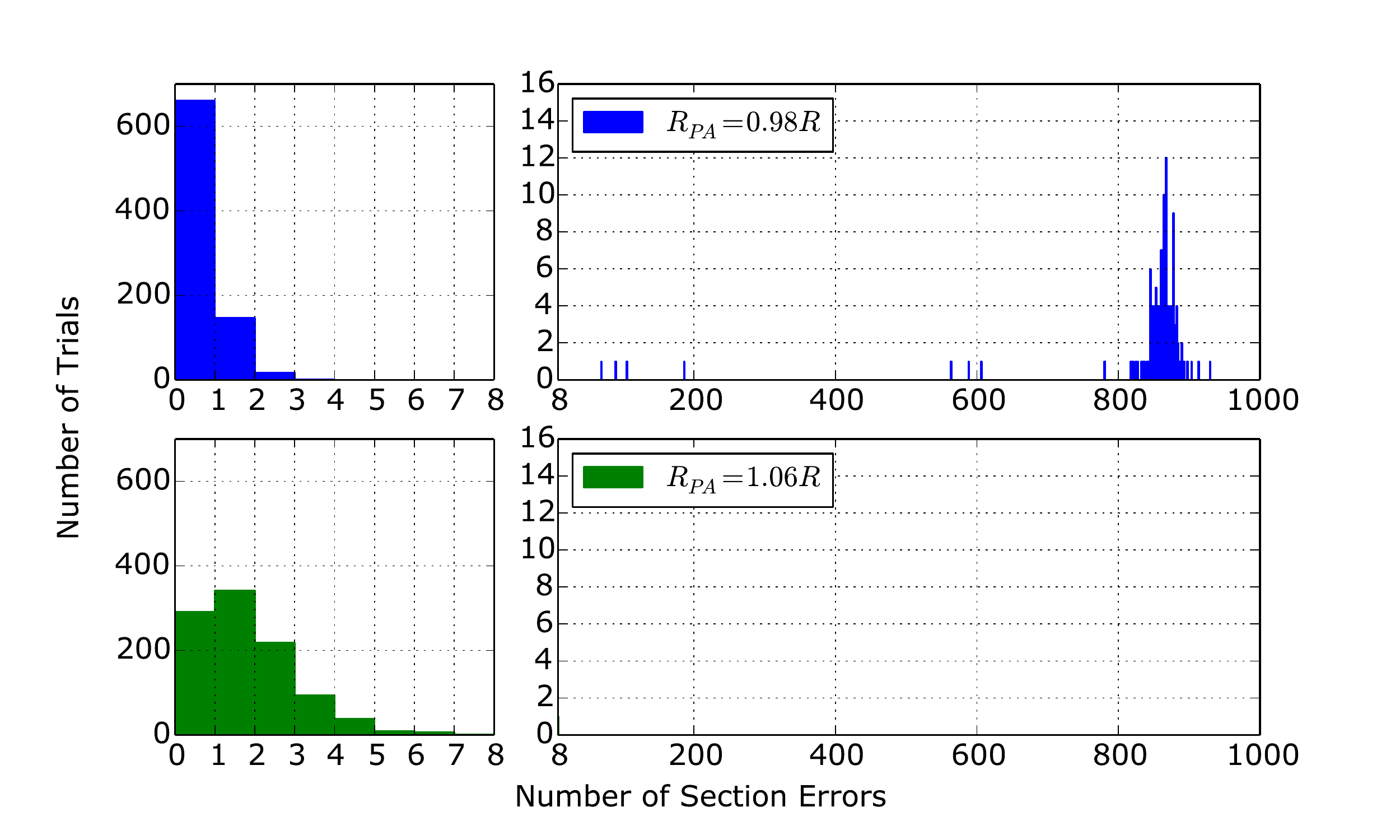}
    \caption{\small Histogram of AMP section errors over $1000$ trials for $R_\text{PA}=0.98R$ (top) and
        $R_\text{PA}=1.06R$ (bottom). The SPARC parameters are $L=1024$, $M=512$, $R=1.6$,
        $\snr=15$.  The left panels highlight distribution of trials with  low section error
        counts (up to 8); the right panels indicate the distribution of infrequent but high-error-count trials. At
        lower $R_\text{PA}$, many more trials have no section errors, but those
        that do often have hundreds. At higher $R_\text{PA}$, at most 7 section
        errors were seen, but many fewer trials had zero section errors.}%
\label{fig:pa_par_comparison}
\vspace{-8pt}
\end{figure}

The non-asymptotic bounds on $x(\tau)$ in Lemma \ref{lem:xlem} indicate that at finite lengths, the minimum power required for a section $\ell$ to decode  in an iteration may be slightly different   than that indicated by the approximation  in \eqref{eq:lemma1b}. 
Recall that the iterative power allocation algorithm in Section \ref{sec:pa:iterative} was designed based on \eqref{eq:lemma1b}.
We can compensate for the difference between the approximation  and the actual value of $x(\tau)$ by running the iterative power allocation in Algorithm
\ref{alg:iterative-pa} using a modified rate $R_{\text{PA}}$ which may be slightly 
different from the  communication rate $R$. The choice of $R_{\text{PA}}$
directly affects the error concentration. We now discuss the mechanism for this
effect and give guidelines for choosing $R_{\text{PA}}$ as a function of $R$.

If we run the power allocation algorithm with $R_\text{PA}> R$,  from
\eqref{eq:lemma1b} we see that additional power is allocated to the initial
blocks, at the cost of less power for the final blocks (where the allocation is
flat). Consequently,  it is less likely that one of the initial sections will decode in
error, but more likely that some number of the later sections will instead.
Figure~\ref{fig:pa_par_comparison} (bottom) shows the effect of choosing a
large $R_\text{PA}=1.06R$: out of a total of $1000$ trials, there were no
trials with more than 7 sections decoded in error (the number of sections
$L=1024$); however, relatively few trials ($29\%$) have zero section errors.

Conversely, choosing $R_\text{PA} < R$ allocates less power to the initial
blocks, and increases the power in the final  sections which have a flat
allocation. This increases the likelihood of the initial section being decoded
in error; in a trial when this happens, there will be a large number of section
errors. However, if the initial sections are decoded correctly, the additional
power in the final sections increases the probability of the trial being
completely error-free.  Thus choosing  $R_{PA}< R$  makes completely error-free
trials more likely, but also increases the likelihood of having trials with a
large number of sections in error. In Figure~\ref{fig:pa_par_comparison} (top),
the smaller $R_\text{PA}=0.98R$ gives zero or one section errors in the
majority ($81\%$) of cases, but the remaining trials typically have a large
number of sections in error.

To summarize, the larger the  $R_{PA}$, the better the concentration of section error rates of individual trials around the overall average.  However, increasing $R_{PA}$ beyond a point just increases the average section error rate because of too little power being allocated to the final sections.

For different values of the communication rate $R$, we  empirically determined an $R_\text{PA}$ that gives the lowest average section error rate, by starting at $R_\text{PA}=R$ and searching  the neighborhood in steps of $0.02R$.
Exceptionally, at low rates  (for $R \leq 1$),  the optimal $R_\text{PA}$ is
found to be $0$, leading to a completely flat power allocation with 
$P_\ell = \frac{P}{L}$ for all $\ell$. We note from \eqref{eq:lemma1b} that for  $1 \geq R >
\frac{P}{ 2 \tau_0^2 \ln 2}$, the large system limit theory does not predict
that we can decode \emph{any} of the  $L$ sections --- this is because no
section is above the threshold in the first iteration of decoding. However, in
practice, we observe that some sections will decode initially (due to the
correct column being aligned favorably with the noise vector), and this
reduces the threshold enough to allow subsequent decoding to continue in most
cases. For $R \leq 1$, when $R_\text{PA}$ closer to $R$ is used, the lower
power in later sections hinders the finite length decoding performance. 

We found that the value of $\frac{R_{PA}}{R}$ that minimizes the average section error rate increases with $R$. In
particular, the optimal  $\frac{R_{PA}}{R}$ was  $0$ for $R \leq 1$; the
optimal $\frac{R_{PA}}{R}$ for $R=1.5$  was close to 1, and for $R=2$, the
optimal $\frac{R_{PA}}{R}$ was between $1.05$ and $1.1$. Though this provides a useful design
guideline, a deeper theoretical analysis of the role of $R_{PA}$ in optimizing
the finite length performance  is an open question.

Finally, a word of caution when empirically optimizing $R_\text{PA}$ to minimize the average section error rate. Due to the loss of concentration as $R_{PA}$ is decreased below $R$,  care must be taken to run  sufficient trials to
ensure that a rare unseen trial with many section errors will not
catastrophically impact the overall average section error rate. For example, in
one scenario with $L=1024, M=512, \snr=15, R=1.4, R_\text{PA}=1.316$, we
observed 192 trials with errors out of 407756 trials, but only 4 of these
trials had more than one error, with between 400 to 600 section errors in those
4 cases. The average section error rate was $\num{5.6e-6}$. With fewer trials,
it is possible that no trials with a large number of section errors would be
observed, leading to an estimated error rate an order of magnitude better, at
around $\num{4.6e-7}$.

\section{Online Computation of $\tau_t^2$ and Early Termination} \label{sec:tau}

Recall that the update step \eqref{eq:eta_update} of the AMP decoder requires
the SE coefficients $\tau_t^2$, for $t \in [T]$. In the standard
implementation \cite{Rush2017}, these coefficients are computed in advance
using the SE equations \eqref{eq:taut_def}--\eqref{eq:xt_tau_def}.  The total number of iterations  $T$  is also determined in advance by computing the number of iterations required  the SE to converge to its fixed point  (to within a specified tolerance). This technique produced effective results, but advance computation is slow as each
of the  $L$ expectations in \eqref{eq:xt_tau_def} needs to be computed
numerically via Monte-Carlo simulation, for each  $t$. A faster approach is to
compute the $\tau_t^2$ coefficients using the asymptotic expression for
$x(\tau)$ given in \eqref{eq:lemma1b}. This gives error performance
nearly identical to the earlier approach with significant time savings, but
still requires advance computation.  Both these methods are referred to as
``offline'' as the $\tau_t^2$ values are computed a priori.

A simple way to estimate $\tau_t^2$ online during the decoding process is as follows. 
In each step $t$, after producing $z^t$ as in \eqref{eq:z_update}, we
estimate \be \wh{\tau}_t^2 = \frac{\norm{z^t}^2}{n} = \frac{1}{n} \sum_{i=1}^n
z_i^2. \ee The justification for this estimate comes from the analysis of the
AMP decoder in \cite{Rush2017,Rush2017b}, which shows that for large $n$,
$\wh{\tau}_t^2$ is close to $\tau_t^2$ in  \eqref{eq:taut_def} with high
probability. In particular, \cite{Rush2017b} provides a concentration
inequality for $\wh{\tau}_t^2$ similar to \eqref{eq:pezero}.  We note that such a similar online estimate has been used previously in various AMP and GAMP algorithms \cite{barbier2014replica,BarbSK2015,BarbKrz15,rangan2011generalized}. Here, we show that in addition to being fast, the online estimator permits an interpretation as a measure of SPARC decoding progress and  provides a flexible termination criterion for the decoder. Furthermore, the error performance with the  online estimator was observed to be the same or slightly better than the offline methods.  


Recall from the discussion at the beginning of
Section \ref{sec:pa} that in each step, we have \be s^t:=\beta^t + A^* z^t
\approx \beta + \tau_t Z, \ee where $Z$ is a standard normal random vector
independent of $\beta$. Starting from $\tau_0^2 = \sigma^2+P$, a judicious
choice of power allocation ensures that  the SE parameter $\tau_t^2$ decreases
with $t$, until it converges at $\tau_T^2 =\sigma^2$ in a finite number of
iterations $T$.

\begin{figure}
    \centering
    \includegraphics[width=0.95\columnwidth]{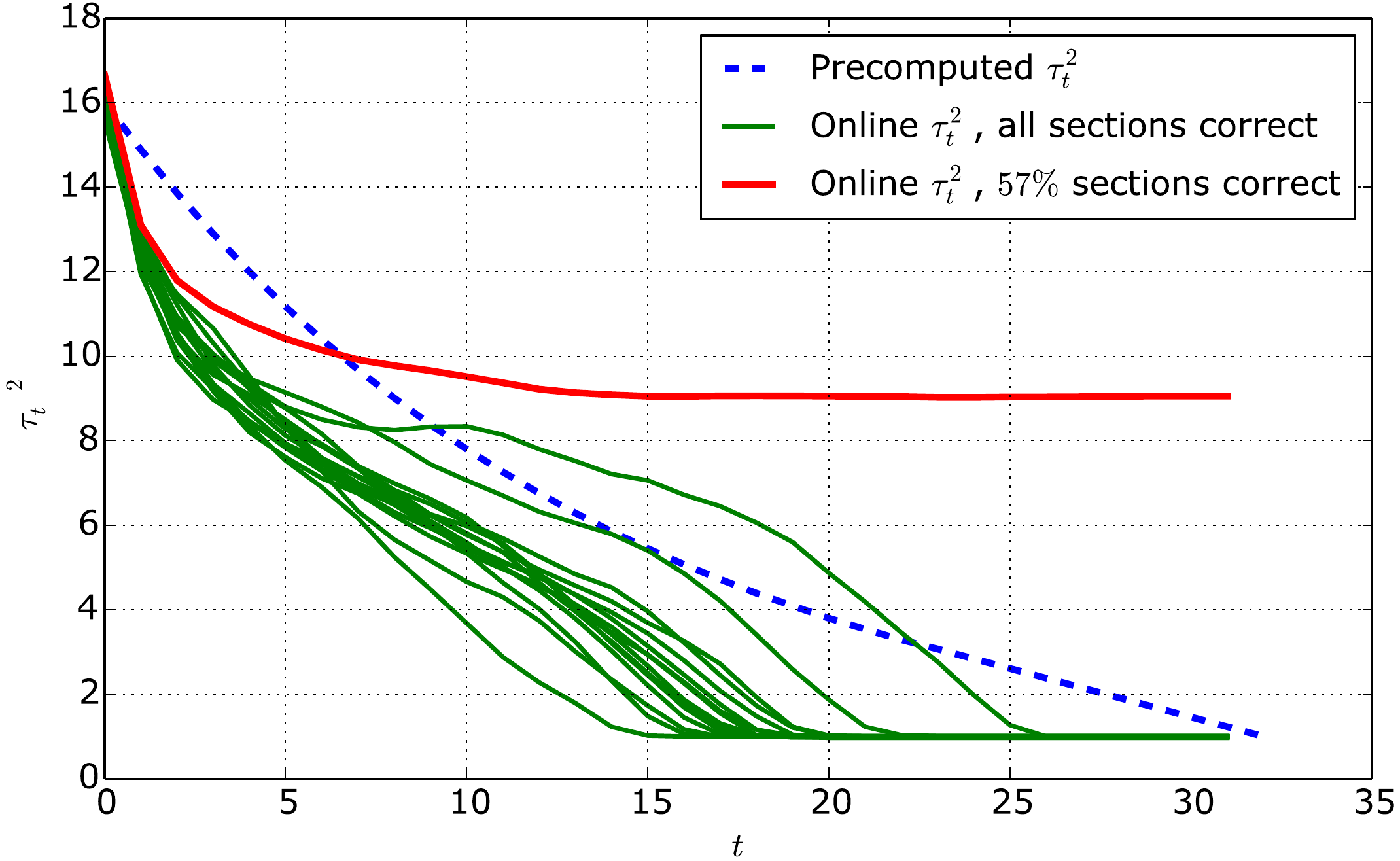}
    \caption{\small Comparison between offline and online  trajectories of the effective noise variance,
              at $L=1024, M=512, P=15, \sigma^2=1, R=1.6$. The dashed line represents the pre-computed SE trajectory of $\tau_t^2$. The plot shows 15 successful runs,  and one uncommon run with many section errors. The true value of $\Var[s^t - \beta]$ during decoding tracks $\wh{\tau}^2_t$ too precisely to distinguish on this plot.}
    \label{fig:tau-comparison-time}
    \vspace{-8pt}
\end{figure}

However, at finite lengths there are deviations from this trajectory of $\tau_t^2$ predicted by SE, i.e., the variance of the effective noise vector $(s^t - \beta)$ may deviate from $\tau_t^2$.  The online estimator $\wh{\tau}_t^2$ is found to  track $\Var(s^t - \beta)=\norm{s^t - \beta}^2/n$ very accurately, even when this variance deviates significantly from $\tau_t^2$.   This effect can be seen in Figure~\ref{fig:tau-comparison-time}, where 16 independent decoder runs are plotted and compared with the SE trajectory for $\tau_t^2$ (dashed line). For the 15 successful runs, the empirical  variance $\Var(s^t - \beta)$ approaches $\sigma^2=1$ along different trajectories depending on how the decoding is progressing. In the unsuccessful run, $\Var(s^t - \beta)$ converges to a value much larger than $\sigma^2$.

 In all the runs,  $\wh{\tau}^2_t$ is indistinguishable from  $\Var(s^t -
 \beta)$. This indicates that we can use the final value $\wh{\tau}_T^2$ to
 accurately estimate the power of the undecoded sections --- and thus the
 number of sections decoded correctly --- at runtime. Indeed, $(\wh{\tau}^2_T -
 \sigma^2)$ is an accurate estimate of the total power in the incorrectly
 decoded sections. This, combined with the fact that the power allocation is
 non-increasing, allows the decoder to estimate the number of incorrectly decoded sections. 

Furthermore, we can use the change in $\wh{\tau}_t^2$ between iterations to terminate the decoder early. If the value $\wh{\tau}_t^2$ has not changed between successive iterations, or the change is within some small threshold, then the decoder has stalled and no further iterations are worthwhile. Empirically we find that a stopping criterion with a small threshold (e.g., stop when $\abs{\wh{\tau}^2_t - \wh{\tau}^2_{t-1}} < P_L$)  leads to no additional errors compared to running the decoder for the full iteration count, while giving a significant speedup in most trials.
Allowing a larger threshold for the stopping criterion  gives even better running time improvements.  This early termination criterion based on $\wh{\tau}_t^2$ gives us flexibility in choosing the number of blocks $B$ in the iterative power allocation algorithm of Section \ref{sec:pa:iterative}. This is because the number of AMP iterations is no longer tied to $B$, hence $B$ can be chosen as large as desired.  

To summarize, the online estimator $\wh{\tau}_t^2$ provides an
estimate of the noise variance in each AMP iteration that accurately reflects how the decoding is progressing  in that trial. It thereby enables the decoder to effectively adapt to  deviations from the $\tau_t^2$ values predicted by SE. This explains the improved performance compared to the offline methods of computing $\tau_t^2$. More importantly, it provides an early termination criterion for the AMP decoder as well as a way to track decoding progress and predict the number of section errors at runtime.

\section{Predicting $\Es$, $\Eb$ and $\Ec$}
\label{sec:fsk}

For a given power allocation $\{ P_\ell \}$ and reasonably large SPARC parameters $(n,M,L)$, it is desirable to have a quick way to estimate the section error rate and codeword error rate, without resorting to simulations. Without loss of generality, we assume that the power allocation is asymptotically good, i.e., the large system limit SE parameters (computed using \eqref{eq:lemma1b}) predict reliable decoding, i.e., the  SE converges to $x_T=1$ and $\tau_T^2 =\sigma^2$ in the large system limit. The goal is to estimate the finite length section error rate $\Es$.

One way to estimate $\Es$ is via the state evolution prediction  \eqref{eq:SE_pred}, using $\tau_T= \sigma$. However, 
computing \eqref{eq:SE_pred} requires computing $L$ expectations,  each involving a function of $M$ independent standard normal random variables.   The following result provides estimates of $\Es$ and $\Ec$ that are as accurate as the SE-based estimates,  but  much simpler to compute.

\begin{prop}
   Let the power allocation $\{ P_\ell \}$ be such that the state evolution iteration using the asymptotic approximation \eqref{eq:lemma1b} converges to $\tau_T^2 =\sigma^2$. Then, under the idealized assumption that $\beta^T+ A^*z^T=\beta + \tau_T Z$ (where $Z$ is a standard normal random vector independent of $\beta$), we have the following. The probability of a section (chosen uniformly at random) being incorrectly decoded is 
   \begin{align}
\bar{\mc{E}}_{\sec} &= 1 -  \frac{1}{L} \sum_{\ell=1}^L  \expec_U \left[  \Phi\left(\frac{\sqrt{nP_\ell}}{\sigma} + U \right)
\right]^{M-1}. \label{eq:est_ser}
\end{align}
The probability of the codeword being incorrectly decoded is
\begin{align}
\bar{\mc{E}}_{\text{cw}} &= 1-\prod_{\ell=1}^L \expec_U \left[
    \Phi\left(\frac{\sqrt{nP_\ell}}{\sigma} + U  \right)
\right]^{M-1}. \label{eq:est_cer}
\end{align}
In both expressions above, $U$ is a  standard normal random variable, and $\Phi(.) $ is the standard normal cumulative distribution function.
\label{prop:serr_quick}
\end{prop}

\begin{IEEEproof}
As $\tau_T^2 = \sigma^2$, the effective observation in the final iteration has the representation $\beta + \sigma Z$. The denoising function $\eta^T$ generates a final estimate based on this effective observation, and the index of the largest entry in each section is chosen to form the  decoded message vector $\widehat{\beta}$. Consider the decoding of section $\ell$ of $\beta$. Without loss of generality, we can assume that the first entry of the section is the non-zero one. Using the notation $\beta_{\ell, j}$ to denote the $j$th entry of the section $\beta_\ell$, we therefore have $\beta_{\ell, 1} = \sqrt{n P_\ell}$, and 
$\beta_{\ell, j} =  0$ for $2 \leq j \leq M$.  As the effective observation for section $\ell$ has the representation 
$(\beta^T+ A^*z^T)_\ell = \beta_\ell +  \sigma Z_\ell$, the section  will be
incorrectly decoded if and only if the following event occurs:
\[ \left\{ \sqrt{nP_\ell} + \sigma Z_{\ell, 1} \leq \sigma Z_{\ell, 2}  \right\}  \cup \ldots \cup  \left\{ \sqrt{nP_\ell} + \sigma Z_{\ell, 1} \leq \sigma Z_{\ell, M}  \right\}. \]
Therefore, the probability that the $\ell$th section is decoded in error can be computed as 
\be
\begin{split}
P_{\text{err},\ell} & = 1 -\Prob \left( \sqrt{nP_\ell} + \sigma Z_{\ell, 1}  >  \sigma Z_{\ell, j}, \ 2 \leq j \leq M   \right) \\
& = 1 -  \int_{\mathbb{R}} \prod_{j=2}^M \Prob \left(  Z_{\ell, j} < \frac{\sqrt{nP_\ell} }{\sigma} + u  \, \Big \vert \,  Z_{\ell, 1}=u \right) \phi(u) du  \\
& = 1 - \expec_U \left[ \Phi \left( \frac{\sqrt{nP_\ell} }{\sigma} + U \right) \right]^{M-1},
\end{split}
\label{eq:Pel}
\ee
where $\phi$ and $\Phi$ denote the density and the cumulative distribution function of the standard normal distribution, respectively.
In the second line of \eqref{eq:Pel},  we condition on $Z_{\ell, 1}$ and  then use the fact that $Z_{\ell,1}, \ldots, Z_{\ell, M}$ are i.i.d. $\sim \mc{N}(0,1)$. 

The  probability of a section chosen uniformly at random being incorrectly decoded is  $\frac{1}{L}\sum_{\ell =1}^L P_{\text{err},\ell}$. The  probability of codeword error is one minus the probability that no section is in error, which is given by $1- \prod_{\ell=1}^L (1 - P_{\text{err},\ell})$.  Substituting for $P_{\text{err},\ell}$ from \eqref{eq:Pel} yields the expressions in \eqref{eq:est_ser} and \eqref{eq:est_cer}.

\end{IEEEproof}

\begin{figure}
    \centering
    \includegraphics[width=0.9\columnwidth]{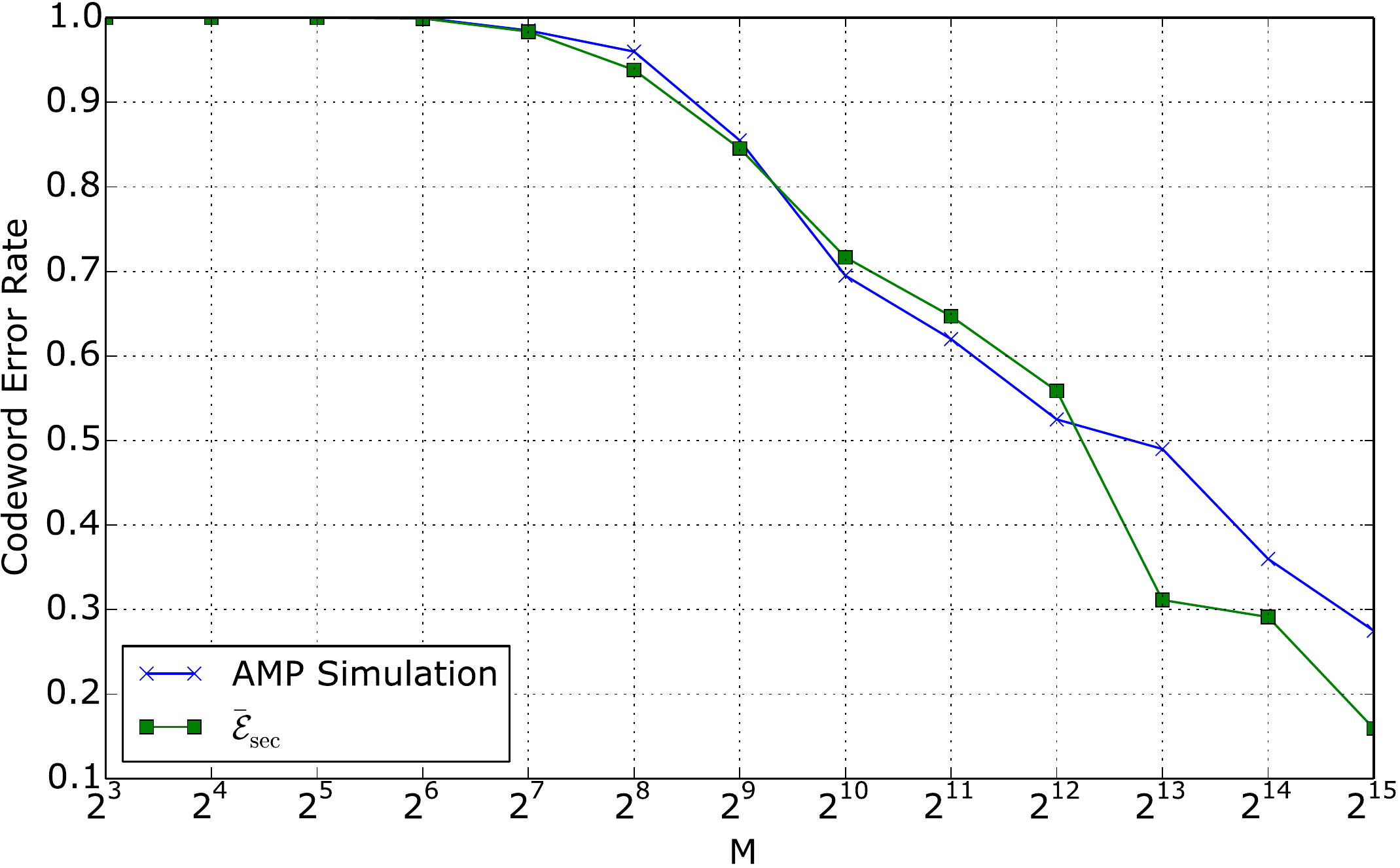}
    \caption{\small Comparison of codeword error rate between simulation results
        and $P_{\text{err}}$-based analysis, for
        $\mc{E}_{\text{cw}}$ with varying $M$.
        $L=1024$, $R=1.5$, $E_b/N_0=5.7$dB. Results are well matched even when
        concentration is poor.}%
\label{fig:fsk-cer}
\vspace{-8pt}
\end{figure}

The section error rate and codeword error rate can be estimated using the idealized expressions in \eqref{eq:est_ser} and \eqref{eq:est_cer}. This still requires computing $L$ expectations, but each expectation is now a function of a single Gaussian random variable, rather than the $M$ independent ones in the SE estimate. Thus we reduce the complexity by a factor of $M$ over the SE approach; evaluations of $\bar{\mc{E}}_{\sec}$ and $\bar{\mc{E}}_{\text{cw}}$ typically complete within a second.

Figure~\ref{fig:lvsm_ser} shows $\bar{\mc{E}}_{\sec}$ alongside the SE estimate $\mc{E}^{\text{SE}}_{\sec}$ for $L=1024$, and various values of $M$.  We see that both these estimates match the simulation results closely up to a certain value of $M$. Beyond this point,  the simulation results diverge from theoretical estimates due to lack of concentration in section error rates across trials, as described in Sec. \ref{sec:lvsm}.  Figure \ref{fig:fsk-cer} compares the idealized codeword error probability in \eqref{eq:est_cer} with that obtained from simulations. Here, there is a good match between the estimate and the simulation results as the concentration of section error rates across trials plays no role --- any trial with one or more section errors corresponds to  one codeword error.


\section{Comparison with Coded Modulation}
\label{sec:ldpc}

\begin{figure}
    \centering
    \includegraphics[width=\columnwidth]{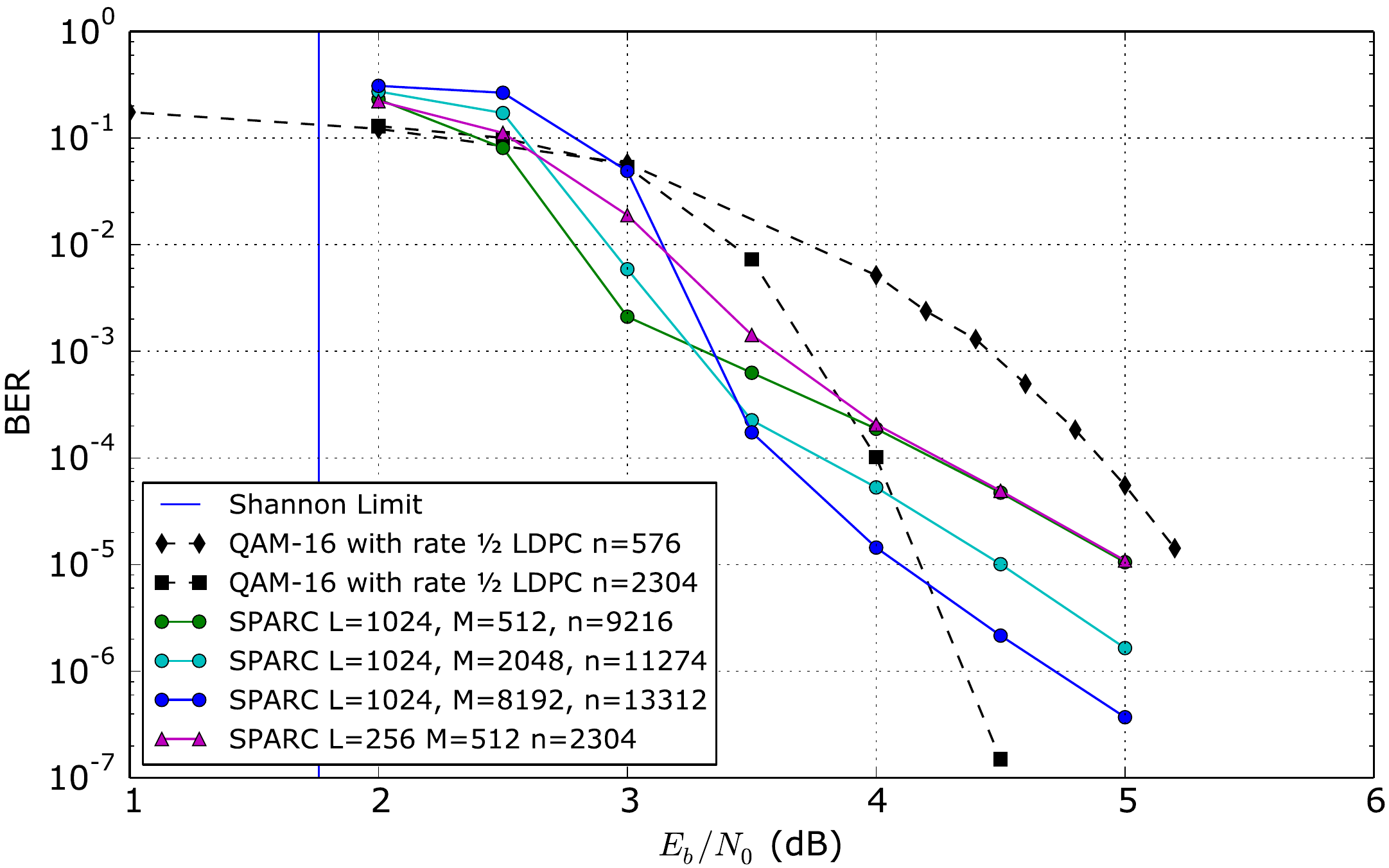}
    \caption{\small Comparison with LDPC coded modulation at $R=1$}
    \label{fig:ldpc-comparison-r10}
\end{figure}
\begin{figure}
    \centering
    \includegraphics[width=\columnwidth]{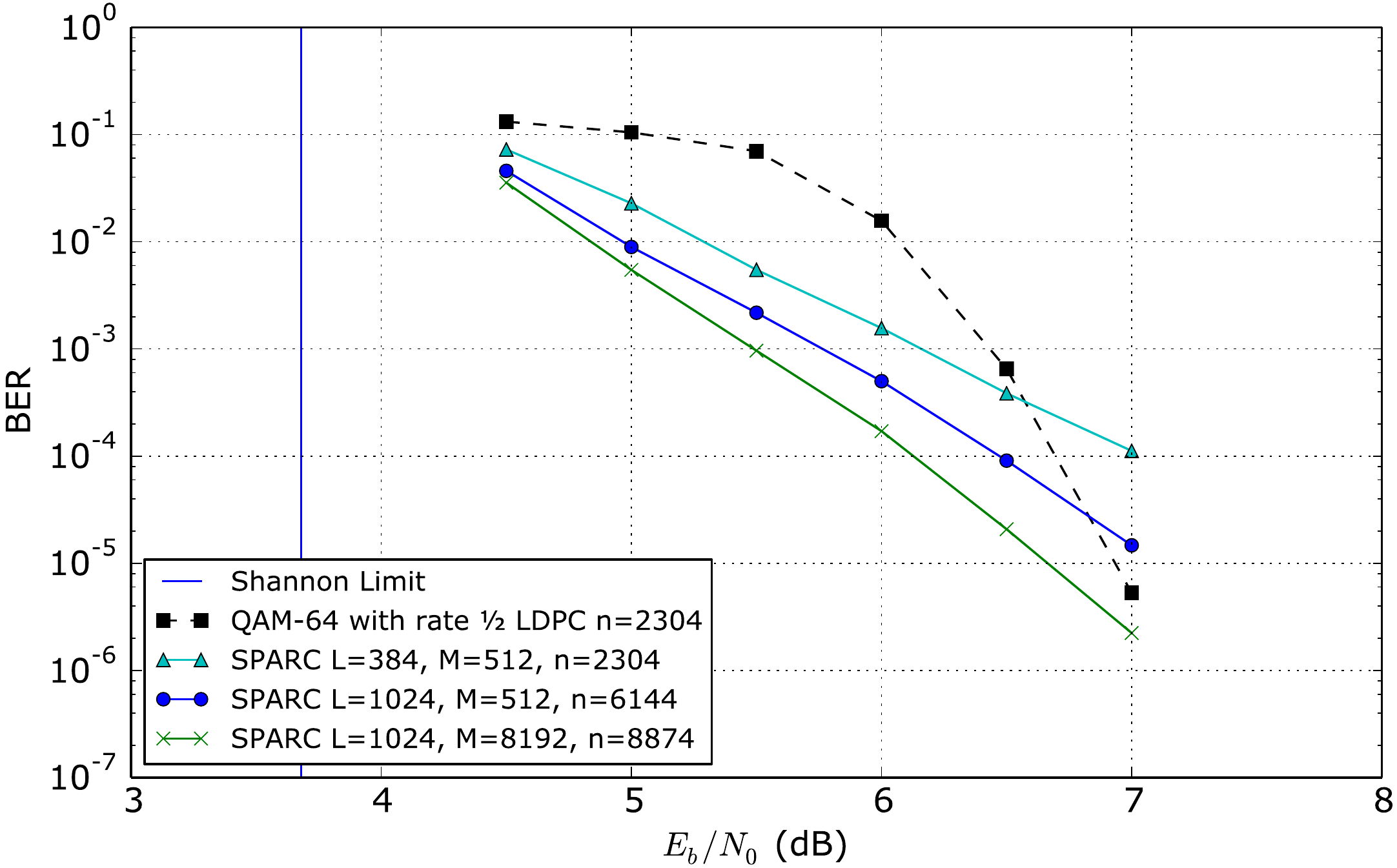}
    \caption{\small Comparison with LDPC coded modulation at $R=1.5$}
    \label{fig:ldpc-comparison-r15}
\end{figure}

In this section, we compare the performance of AMP-decoded SPARCs against coded modulation with LDPC codes.   Specifically, we compare with two instances of coded modulation with LDPC codes from the WiMax standard IEEE 802.16e: 1)  A $16$-QAM constellation with a rate $\frac{1}{2}$ LDPC code  for an overall rate  $R= 1$ bit/channel use/real dimension, and 2) A $64$-QAM constellation with a rate $\frac{1}{2}$ LDPC code  for an overall rate $R= 1.5$ bits/channel use/real dimension. (The spectral efficiency is $2R$ bits/s/Hz.)
The coded modulation results, shown in dashed lines in Figures \ref{fig:ldpc-comparison-r10} and \ref{fig:ldpc-comparison-r15}, are obtained using the CML toolkit \cite{cml} with LDPC code lengths $n=576$ and $n=2304$.

Each figure compares the bit error rates (BER) of the coded modulation schemes with various SPARCs of the same rate, including a SPARC with a matching code length of $n=2304$.  Using $P=E_b R$ and $\sigma^2 = \frac{N_0}{2}$,
 the signal-to-noise ratio of the SPARC can be expressed as  $\frac{P}{\sigma^2} = \frac{2R E_b}{N_0}$.   The SPARCs are implemented using Hadamard-based design matrices,  power allocation designed using the iterative algorithm in Sec. \ref{sec:pa:iterative} with $B=L$, and online $\wh{\tau}_t^2$ parameters with the early termination criterion (Sec. \ref{sec:tau}).  An IPython notebook detailing the SPARC implementation is available at \cite{adamSPARC}.

Figure \ref{fig:ldpc-comparison-r10} shows that for $L=1024$, the best value of $M$ among those considered increases from $M=2^{9}$ at lower $\snr$ values to $M=2^{13}$ at higher $\snr$ values. This is due to the effect discussed in Section~\ref{sec:lvsm}, where larger $\snr$ values can support larger values of $M$, before performance starts degrading due to loss of concentration.

At both $R=1$ and $R=1.5$, the SPARCs outperform the LDPC coded modulation at  $E_b/N_0$ values close to the Shannon limit, but the error rate
does not drop off as quickly at higher values of $E_b/N_0$.   One way to enhance SPARC performance at higher $\snr$  is by treating it as a high-dimensional modulation scheme and adding an outer code.  This is the focus of the next section.


\section{AMP with Partial Outer Codes}
\label{sec:ldpc-outer}

Figures~\ref{fig:ldpc-comparison-r10} and~\ref{fig:ldpc-comparison-r15} show
that for block lengths of the order of a few thousands, AMP-decoded SPARCs
do not exhibit a steep waterfall in section error rate.  Even at high $E_b/N_0$ values, it is still common to observe a small
number of section errors. If these could be corrected, we could hope to obtain
a sharp waterfall behavior similar to the LDPC codes.

 In the simulations of the AMP decoder described above, when $M$
and $R_\text{PA}$ are chosen such that the average error rates are well-concentrated around the state evolution prediction, the number of section errors observed is similar across
 trials.  Furthermore, we observe that the majority
of sections decoded incorrectly are those in the flat region of the power
allocation, i.e., those with the lowest allocated power. This suggests we could use a
high-rate outer code to protect just these sections, sacrificing some rate, but
less than if we na\"{\i}vely protected all sections. We call the sections covered by the outer code \emph{protected} sections, and conversely the earlier sections which are not covered by the outer code  are  \emph{unprotected}.   In \cite{Antony2012}, it was shown that  a Reed-Solomon outer code (that covered all the sections)  could be used to obtain a bound the probability of codeword error from  a bound on the probability of excess section error rate.

Encoding with  an outer code (e.g., LDPC or Reed-Solomon code) is straightforward:  just replace the message bits corresponding to the protected sections with coded bits generated using the usual encoder for the chosen outer code. To decode, we would like to obtain bit-wise posterior probabilities for each  codeword bit of the outer
code, and use them as inputs to a soft-information decoder, such as a
sum-product or min-sum decoder for LDPC codes. The output of the AMP decoding
algorithm permits this: it yields $\beta^T$, which contains weighted \emph{section-wise} posterior
probabilities; we can directly transform these into \emph{bit-wise}
posterior probabilities. See
Algorithm~\ref{alg:pp2bp} for details.

Moreover, in addition to correcting AMP decoding errors in the protected sections, successfully decoding the outer code also provides a way to correct  remaining errors in the unprotected sections of the SPARC codeword. Indeed, after decoding the outer code we can subtract the contribution of the protected sections from the channel output sequence 
$y$, and  re-run the AMP decoder on just the unprotected sections. The key point  is that subtracting the contribution of the later (protected) sections eliminates the interference due to these sections; then running  the AMP decoder on the unprotected sections is akin to operating at a much lower rate.

Thus the decoding procedure has three stages:  i)  first round of AMP  decoding, ii)  decoding the outer code using soft outputs from the AMP, and  iii)  subtracting the contribution of the sections protected by the outer code, and running the AMP decoder again for the unprotected sections. We find that the final stage, i.e., running the AMP decoder again after the outer code recovers errors in the protected sections of the SPARC, provides a significant advantage over a standard application of
an outer code, i.e., decoding the final codeword after the second stage.

We describe this combination of SPARCs with outer codes below, using an
LDPC outer code. The resulting error rate curves exhibit sharp waterfalls in
final error rates, even when the LDPC code
only covers a minority of the SPARC sections.

\begin{figure}[t]
    \centering
    \includegraphics[width=\columnwidth]{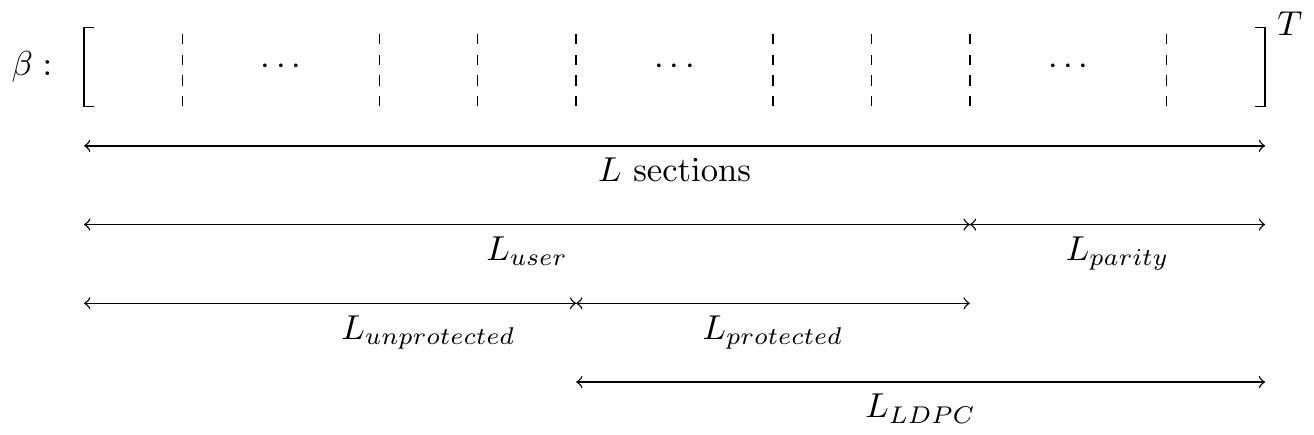}
    \caption{\small Division of the $L$ sections of $\beta$ for an outer LDPC code}
    \label{fig:ldpc-outer-code-division}
    \vspace{-8pt}
\end{figure}

We use a binary LDPC outer code with rate  $R_{LDPC}$, block
length $n_{LDPC}$ and code dimension $k_{LDPC}$, so that
$k_{LDPC}/n_{LDPC}=R_{LDPC}$. For clarity of exposition we assume that both
$n_{LDPC}$ and $k_{LDPC}$ are multiples of $\log M$ (and consequently that $M$
is a power of two). As each section of the SPARC corresponds to $\log M$ bits, if $\log M$ is an integer, then $n_{LDPC}$ and $k_{LDPC}$ bits represent an
integer number of SPARC sections, denoted by  $$L_{LDPC} = \frac{n_{LDPC}}{\log M}  \quad \text{and} \quad L_{protected} = \frac{k_{LDPC}}{\log M},$$ respectively. The assumption that $k_{LDPC}$ and  $n_{LDPC}$ are multiples of $\log M$ is not necessary in practice; the general case is discussed at the end of the next subsection.

We partition the $L$  sections of the SPARC codeword as shown in Fig~\ref{fig:ldpc-outer-code-division}. There are $L_{user}$ sections corresponding to the user (information) bits; these sections are divided into
\emph{unprotected} and \emph{protected} sections, with only the latter being covered by the outer LDPC code.  The parity bits of the LDPC codeword index the last $L_{parity}$ sections of the SPARC. For convenience, the \emph{protected} sections and the  \emph{parity} sections  together are referred to as the  \emph{LDPC} sections.

For a numerical example, consider the case where $L=1024$, $M=256$. There are
$\log M=8$ bits per SPARC section. For a $(5120, 4096)$ LDPC code ($R_{LDPC}=4/5$)
we obtain the following relationships between the number of the sections of each kind:
\begin{align*}
  &   L_{parity}=\frac{n_{LDPC}-k_{LDPC}}{\log M}=\frac{(5120-4096)}{8}=128,\\
    &     L_{user} =L-L_{parity}=1024-128=896, \\
  &   L_{protected}=\frac{k_{LDPC}}{\log M}= \frac{4096}{8}=512,\\
   & L_{LDPC} =L_{protected} +L_{parity}=512+128=640, \\
      &     L_{unprotected} = L_{user} -  L_{protected} = L-L_{LDPC}=384. 
\end{align*}
There are $L_{user}\log M=7168$ user bits, of which the final $k_{LDPC}=4096$
are encoded to a systematic $n_{LDPC}=5120$-bit LDPC codeword. The resulting
$L\log M=8192$ bits (including both the user bits and the LDPC parity bits) are
encoded to a SPARC codeword using the  SPARC encoder and power allocation described in previous sections.

We continue to use $R$ to denote the overall user rate, and $n$ to denote the SPARC code length so that  $nR=L_{user}\log M$.
The  underlying SPARC rate (including the overhead due to the outer code) is denoted by $R_{SPARC}$. We note that 
$nR_{SPARC}=L\log M$, hence $R_{SPARC} > R$. For example,  with $R=1$ and $L,M$ and the outer code parameters as chosen above, $n=L_{user}(\log M)/R=7168$, so $R_{SPARC}=1.143$.

\begin{algorithm}[t]
    \caption{Weighted position posteriors $\beta_\ell$ to bit posteriors $p_0,\ldots,p_{\log M - 1}$ for section $\ell\in [L]$}
    \label{alg:pp2bp}
    \begin{algorithmic}
        \REQUIRE $\beta_\ell=[\beta_{\ell,1},\ldots,\beta_{\ell,M}]$, for $M$ a power of 2
        \STATE Initialise bit posteriors $p_0,\ldots,p_{\log M -1} \leftarrow 0$
        \STATE Initialise normalization constant $c \leftarrow \sum_{i=1}^{M}\beta_{\ell,i}$
        \FOR{$\log i=0,1,\ldots,\log M - 1$}
            \STATE $b \leftarrow \log M - \log i - 1$
            \STATE $k \leftarrow i$
            \WHILE{$k < M$}
                \FOR{$j=k+1, k+2, \ldots, k+i$}
                    \STATE $p_b \leftarrow p_b + \beta_{\ell,j}/c$
                \ENDFOR
                \STATE $k \leftarrow k + 2i$
            \ENDWHILE
        \ENDFOR
        \RETURN $p_0,\ldots,p_{\log M -1}$
    \end{algorithmic}
\end{algorithm}

\subsection{Decoding SPARCs with LDPC outer codes}

At the receiver, we decode as follows:

\begin{enumerate}
    \item Run the AMP decoder to obtain $\beta^T$. Recall that entry $j$ within section $\ell$ of $\beta^T$
is proportional to  the posterior probability of the column $j$ being the
transmitted one for section $\ell$. Thus the AMP decoder gives section-wise posterior probabilities for each section $\ell \in [L]$. 

    \item Convert the section-wise posterior probabilities to bit-wise posterior probabilities using
Algorithm~\ref{alg:pp2bp}, for each of the $L_{LDPC}$ sections.  This requires
$O(L_{LDPC}M \log M)$ time complexity, of the same order as one iteration of
AMP.

    \item Run the LDPC decoder using the bit-wise posterior probabilities obtained in Step 2 as inputs.

    \item If the LDPC decoder fails to produce a valid LDPC codeword, terminate decoding here,
using $\beta^T$  to produce $\hat{\beta}$ by selecting the maximum value in each
section (as per usual AMP decoding).

    \item If the LDPC decoder succeeds in finding a  valid codeword, we use it to re-run AMP decoding on
        the unprotected sections. For this, first convert the LDPC codeword bits to a partial
        $\hat{\beta}_{LDPC}$ as follows, using a method similar to the original SPARC encoding:
        \begin{enumerate}
            \item Set the first $L_{unprotected}M$ entries of  $\hat{\beta}_{LDPC}$  to zero,
            \item The remaining $L_{LDPC}$ sections (with $M$ entries per section) of $\hat{\beta}_{LDPC}$ will have exactly one-non zero entry per section, with the LDPC codeword   determining the location of the non-zero in each section.     Indeed, noting that $n_{LDPC}= L_{LDPC} \log M$, we consider the LDPC codeword as a concatenation of $L_{LDPC}$ blocks of $\log M$ bits each, so that each block of bits indexes the location of the non-zero entry in one section of  $\hat{\beta}_{LDPC}$.  The value of the non-zero in section $\ell$ is set to  $\sqrt{n P_\ell}$, as per the power allocation.
        \end{enumerate}
        Now subtract the codeword corresponding to $\hat{\beta}_{LDPC}$ from
        the original channel output $y$, to obtain $y'=y-A\hat{\beta}_{LDPC}$.

    \item Run the AMP decoder again, with input $y'$, and operating only
        over the first $L_{unprotected}$ sections. As this operation is effectively at a much
        lower rate than the first decoder (since the interference contribution from
       all the protected sections is removed), it is more likely that the
        unprotected bits are decoded correctly than in the first AMP decoder.
        
     We note that instead of generating $y'$, one could run the AMP decoder directly on $y$, but enforcing that in each AMP iteration, each  of the $L_{LDPC}$ sections has all its non-zero mass on the entry determined by $\hat{\beta}_{LDPC}$, i.e., consistent with Step 5.b).

    \item Finish decoding, using the output of the final AMP decoder to find
        the first $L_{unprotected}M$ elements of $\hat{\beta}$, and using
        $\hat{\beta}_{LDPC}$ for the remaining $L_{LDPC}M$ elements.

\end{enumerate}

In the case where $n_{LDPC}$ and $k_{LDPC}$ are not multiples of $\log M$, the values $L_{LDPC} =n_{LDPC}/ \log M$ and  $L_{protected} = k_{LDPC}/ \log M$
will not be integers. Therefore one section at the boundary of $L_{unprotected}$ and $L_{protected}$
will consist of some unprotected bits and some protected bits. Encoding is not affected in this situation, as the LDPC encoding happens prior
to SPARC codeword encoding. When decoding, conversion to bit-wise posterior probabilities is performed for all
sections containing LDPC bits (including the intermediate section at the
boundary) and only the $n_{LDPC}$ bit posteriors corresponding to the LDPC
codeword are given to the LDPC decoder. When forming $\hat{\beta}_{LDPC}$, the
simplest option is to treat the intermediate section as though it were
unprotected and set it to zero. It is also possible to compute column
posterior probabilities which correspond to the fixed LDPC bits and
probabilities arising from $y$, though doing so is not covered in this paper.

\subsection{Simulation results}

The combined AMP and outer LDPC setup described above was simulated using the
(5120, 4096) LDPC code ($R_{LDPC}=4/5$) specified in \cite{CCSDS131} with a
min-sum decoder. Bit error rates were measured only over the user bits, ignoring any bit
errors in the LDPC parity bits.

\begin{figure}
    \centering
    \includegraphics[width=\columnwidth]{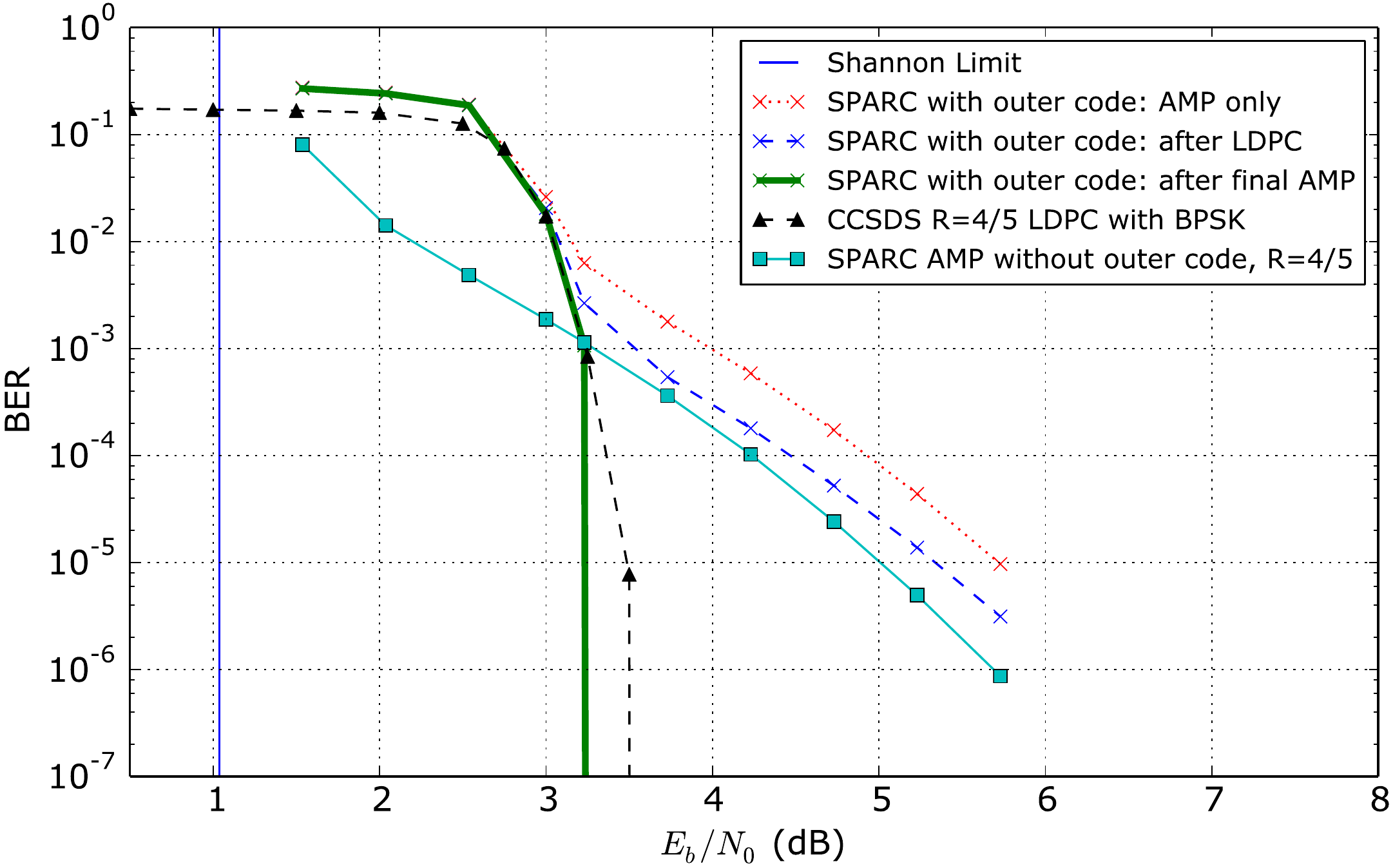}
    \caption{\small Comparison to plain AMP and to BPSK-modulated LDPC at overall rate $R=0.8$.
    The SPARCs are both $L=768$, $M=512$. The underlying SPARC rate when the outer code is included is $R_{SPARC}=0.94$. The BPSK-modulated LDPC is the same CCSDS LDPC
code \cite{CCSDS131} used for the outer code. For this configuration, $L_{user}=654.2$, $L_{parity}=113.8$,
$L_{unprotected}=199.1$, $L_{protected}=455.1$, and $L_{LDPC}=568.9$.}
    \label{fig:ldpc-outer-r08}
\end{figure}

\begin{figure}
    \centering
    \includegraphics[width=\columnwidth]{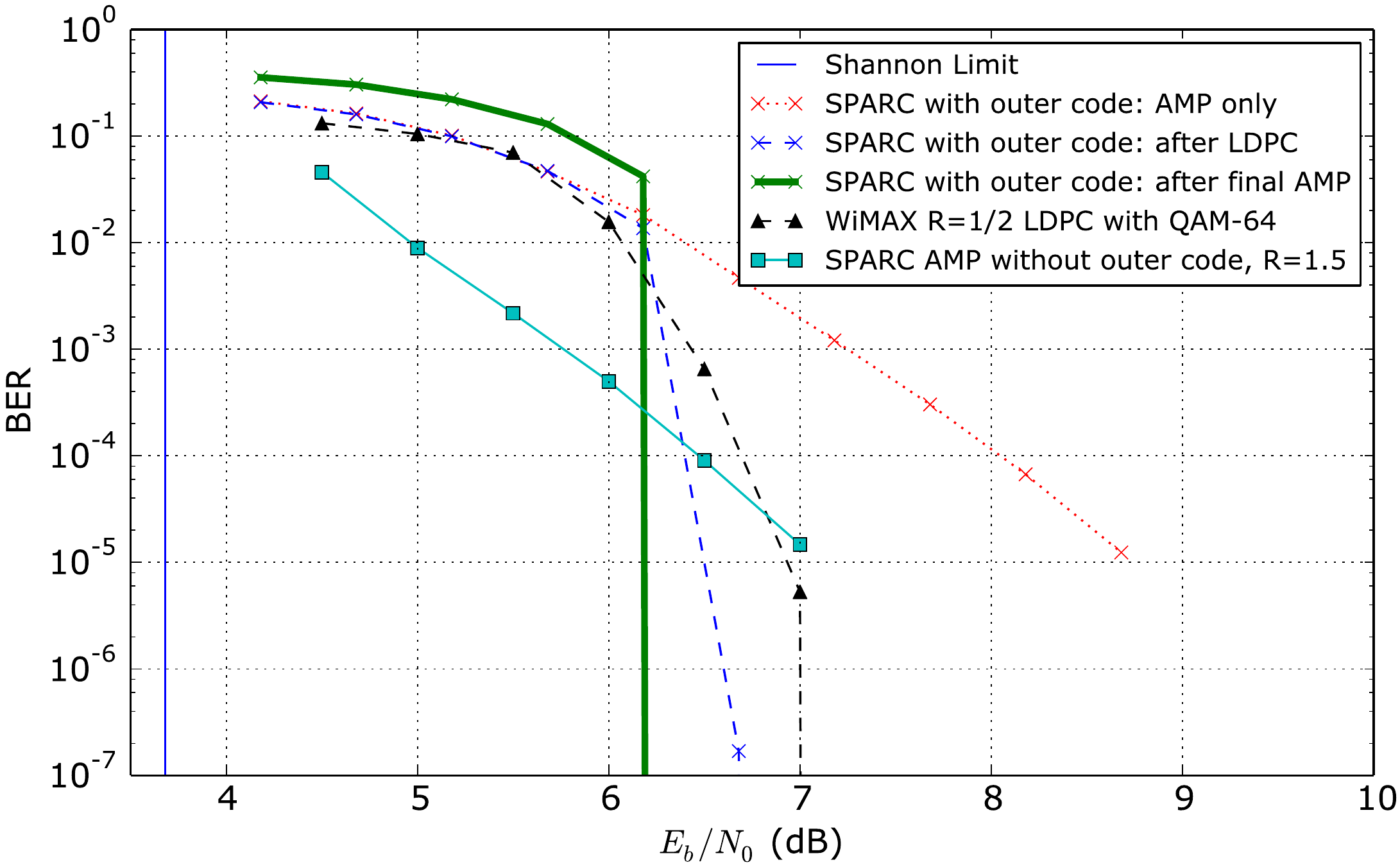}
    \caption{\small Comparison to plain AMP and to the QAM-64 WiMAX LDPC of Section~\ref{sec:ldpc} at overall rate $R=1.5$
    The SPARCs are both $L=1024$, $M=512$. The underlying SPARC rate including the
outer code is $R_{SPARC}=1.69$. For this configuration, $L_{user}=910.2$,
$L_{parity}=113.8$, $L_{unprotected}=455.1$, $L_{protected}=455.1$, and $L_{LDPC}=455.1$.}
    \label{fig:ldpc-outer-r15}
    \vspace{-8pt}
\end{figure}

Figure~\ref{fig:ldpc-outer-r08} plots results at overall rate $R=\frac{4}{5}$, where
the underlying LDPC code (modulated with BPSK) can be compared to the 
SPARC with LDPC outer code, and to a plain SPARC  with rate $\frac{4}{5}$. In this case
$R_{PA}=0$, giving a flat power allocation. Figure~\ref{fig:ldpc-outer-r15} plots results at overall rate $R=1.5$, where
we can compare to the QAM-64 WiMAX LDPC code, and to the plain SPARC with rate 1.5  of
Figure~\ref{fig:ldpc-comparison-r15}.

The plots show that  protecting a fraction of sections with an outer code does provide a steep waterfall above a threshold value of $\frac{E_b}{N_0}$. Below this threshold, the combined SPARC + outer code has worse performance than the plain rate $R$ SPARC without the outer code. This  can be explained as follows. The combined code has a higher SPARC rate
$R_{SPARC}>R$, which leads to  a  larger section error rate for the first AMP decoder, and consequently,  to worse bit-wise posteriors at the input of the LDPC decoder. For $\frac{E_b}{N_0}$ below the threshold, the noise level at the input of the LDPC decoder is beyond than the error-correcting capability of the LDPC code, so the LDPC code effectively does not correct any section errors. Therefore the overall performance is worse than the performance without the outer code.

Above the threshold, we observe that the second AMP decoder (after subtracting the contribution of the LDPC-protected sections) is successful at
decoding the unprotected sections that were initially decoded incorrectly. This
is especially apparent in the  $R=\frac{4}{5}$ case (Figure \ref{fig:ldpc-outer-r08}), where the section errors are
uniformly distributed over all sections due to the flat power allocation;
errors are just as likely in the unprotected sections as in the protected
sections.

\subsection{Outer code design choices}

In addition to the various SPARC parameters discussed in previous sections,
performance with an outer code is sensitive to what fraction of sections are
protected by the outer code. When more sections are protected by the outer code, the overhead of using the outer code is also
higher, driving $R_{SPARC}$ higher for the same overall user rate $R$. This
leads to worse performance in the initial AMP decoder, which has to operate
at the higher rate $R_{SPARC}$. As discussed above, if $R_{SPARC}$ is increased too much, the bit-wise 
posteriors input to the LDPC decoder are degraded beyond its ability to successfully decode,
giving poor overall performance.

Since the number of sections covered by the outer code depends on both $\log M$ and $n_{LDPC}$,
various trade-offs are possible. For example,  given $n_{LDPC}$, choosing a larger value of $\log M$ corresponds to  fewer sections being covered by the outer code. This results in smaller rate overhead, but increasing $\log M$ may also affect concentration of the error rates around the SE predictions, as discussed in Section~\ref{sec:lvsm}. We conclude with two remarks about the choice  of parameters for the SPARC and the outer code. 
\begin{enumerate}
\item When using an outer code, it is highly beneficial to have good concentration of
the section error rates for the initial AMP decoder. This is because a small
number of errors in a single trial can usually be fully corrected by the outer code, while
occasional trials with a very large number of errors cannot.

\item Due to the second AMP decoder operation, it is not necessary for all sections
with low power to be protected by the outer code. For
example,  in Figure~\ref{fig:ldpc-outer-r08}, all sections have equal power, and around 30\% are not protected by the outer code. Consequently, these sections
are  often not decoded correctly by the first decoder. Only once
the protected sections are removed is the second decoder  able to correctly
decode these unprotected sections. In general the aim should be to cover all or most of the sections in the flat
region of the power allocation, but experimentation is necessary to determine
the best trade-off.
\end{enumerate}
An interesting direction for future work would be to develop an EXIT chart  analysis to jointly optimize the design of the SPARC and the outer LDPC code.

\section*{Acknowledgement}
The authors thank the Editor and the anonymous referees for several helpful comments which improved the paper.



\end{document}